\documentclass[aps, prd, nofootinbib,
    amsmath, a4paper, 10pt, twocolumn, 
    showkeys, superscriptaddress,preprintnumbers]{revtex4-2}
\usepackage[english]{babel}
\usepackage[margin=2cm]{geometry}
\usepackage{listings}
\usepackage{graphicx}
\usepackage{amssymb, amsmath}
\usepackage{float}
\usepackage{hyperref}
\usepackage[dvipsnames]{xcolor}
\definecolor{dourado}{rgb}{0.56, 0.45, 0.0}

\begin{document}

\preprint{MIT-CTP/6039}

\title{\bf Reheating matters: Starobinsky inflation in light of joint \\ CMB+BAO results and gravitational-wave forecasts}

\author{Gl\'auber C. \surname{Dorsch}}\thanks{glauber@fisica.ufmg.br}
\affiliation{Departamento de F\'isica, Universidade Federal de Minas Gerais (UFMG), Belo Horizonte, MG, Brazil}
\author{Lucas P. C. \surname{Leão}}\thanks{lucaspatric@ufmg.br}
\affiliation{Departamento de F\'isica, Universidade Federal de Minas Gerais (UFMG), Belo Horizonte, MG, Brazil}
\author{Lucas \surname{M. B. Alves}}\thanks{lmba@mit.edu}
\affiliation{Center for Theoretical Physics -- A Leinweber Institute, Massachusetts Institute of Technology, Cambridge, MA 02139, USA}
\author{Luiz C. A. \surname{Miranda}}\thanks{luizcarlos.miranda@hotmail.com}
\affiliation{Departamento de F\'isica, Universidade Federal de Minas Gerais (UFMG), Belo Horizonte, MG, Brazil}
\author{Beatriz M. D. \surname{Sena}}\thanks{beatrizmotta@ufmg.br}
\affiliation{Departamento de F\'isica, Universidade Federal de Minas Gerais (UFMG), Belo Horizonte, MG, Brazil}

\begin{abstract}
    It has been noted in the literature that, if post-inflationary reheating is dominated by a stiff fluid with equation of state (EoS) $p>\rho/3$, then the predictions of Starobinsky inflation for the scalar spectral index could be made to agree with measurements from the combined CMB+BAO datasets performed by the Atacama Cosmology Telescope (ACT) and the South Pole Telescope (SPT) collaborations. However, a side-effect of such a stiff epoch is the blue-tilting of the primordial gravitational-wave (GW) spectrum. In this work, we explore the observational consequences of this blue-tilting in three scenarios: (i) a purely stiff-dominated reheating, (ii) a more realistic case where reheating is first dominated by a matter-like fluid (corresponding to inflaton oscillations around the bottom of a quadratic potential well) later followed by a stiff epoch, and (iii) a case analogous to the previous one, but with an earlier radiation-dominated instead of matter-dominated epoch. We show that in all cases the $1\sigma$ region allowed by recent CMB+BAO data is already excluded by constraints on the amount of radiation present during Big Bang Nucleosynthesis (BBN). Moreover, in a considerable fraction of the remaining $2\sigma$ region we find that the blue-tilting would be severe enough to make the primordial spectrum detectable in future interferometers such as Einstein Telescope, LISA, DECIGO, and BBO, thus rendering these scenarios testable by these experiments.
\end{abstract}
\maketitle

\section{Introduction}
In the past couple of decades, measurements of the cosmic microwave background (CMB) and baryon acoustic oscillations (BAO) have reached an extraordinary degree of precision, ushering us into what can be dubbed an age of \emph{precision cosmology}, allowing us to constrain or even definitely exclude many models of primordial cosmic inflation. This astounding achievement---being able to make inferences about the very first moments of our cosmos---is a merit both of the experiments and of the theory, since it relies on inflation's power to predict the spectral shape of anisotropies arising from quantum fluctuations during the accelerated expansion of spacetime. In particular, these models typically predict with reasonable accuracy the spectral tilt of scalar perturbations $n_s$ and the tensor-to-scalar ratio $r$ (the ratio of power spectrum amplitudes for tensor and scalar perturbations), both quantities assessed experimentally to high precision. 

Indeed, for over a decade the standard baseline for these constraints has been provided by the Planck satellite~\cite{Planck:2018vyg}, which (combined with BICEP2/Keck Array data~\cite{BICEP:2021xfz}) placed stringent bounds on the primordial scalar spectral index,  $n_s = 0.9665 \pm 0.0038$ (68\% C.L.), and an upper bound on the tensor-to-scalar ratio of $r<0.036$ (95\% C.L.). These results strongly disfavoured many proposed models for slow-roll inflation while strongly favouring Starobinsky's model, in which inflation is seen as a purely gravitational effect due to a modification of the Einstein-Hilbert action by the addition of an $R^2$ term~\cite{Starobinsky:1980te, Starobinsky:1983zz}. Such a term is expected to emerge from quantum corrections to gravity~\cite{tHooft:1974toh, Donoghue:1994dn} and suffices to explain the dynamics of this inflationary epoch, thus dispensing with the \emph{ad hoc} inclusion of any additional inflaton fields. Moreover, it has recently been shown that the model can also successfully account for post-inflationary reheating from purely gravitational interactions, without the need to include further \emph{ad hoc} interactions with the Standard Model~\cite{Dorsch:2024nan, Dorsch:2026ref}. As for the spectral index, the model predicts a relation $n_s\approx1-2/N_*$~\cite{DeFelice:2010aj}, where $N_*$ is the number of $e$-folds of exponential expansion between the first crossing of the Hubble sphere by the CMB pivot scale ($k_*=0.05~\text{Mpc}^{-1}$) and the later time when inflation ends. If one assumes a standard cosmological evolution, with the reheating period dominated by a matter-like fluid, and a reheating temperature at least above the electroweak scale, one finds typical values of $N_*=50-60$~\cite{Liddle:2003as, Dodelson:2003vq}, which lead to $n_s = 0.960 - 0.967$, well within the Planck result\footnote{It is worth noting that ref.~\cite{Liddle:2003as} had already pointed to the possibility of increasing $N_*$ via an early stiff-dominated period, a case which we explore here.}.

However, the whole situation took a sharp turn after the recent DR6 data release from the Atacama Cosmology Telescope (ACT)~\cite{ACTDR6_cosmo_param}---an earth-based telescope measuring small-scale CMB fluctuations---as well as the newest data release on galaxy BAO data from the Dark Energy Spectroscopic Instrument (DESI) collaboration~\cite{DESIDR2}. Their latest joint analysis, combining these cutting-edge probes into the comprehensive P-ACT-LB2-BK18 dataset (Planck + ACT DR6 + DESI DR2 + BICEP/Keck), yields a significantly higher value of $n_s = 0.9752 \pm 0.0030$ at the 68\% confidence level (and $r<0.038$ at 95\% C.L.)~\cite{ACTDR6}. A similar shift of $n_s$ has also been verified by the combined analysis of the South Pole Telescope (SPT) and DESI DR2 data~\cite{SPT-3G:2025bzu}. This updated measurement represents a significant deviation from the Planck legacy results, and reverses the prevailing scenario described above, now putting Starobinsky's model at a $\sim 2\sigma$ tension with the data~\cite{ACTDR6}.

At this point, it is important to emphasize that the CMB data from ACT, SPT, and BICEP/Keck, taken by themselves, are actually fully consistent with Planck measurements and hence also with the predictions of Starobinsky inflation. It is only after including the DESI DR2 BAO dataset that the spectral index is sharply shifted towards larger values, giving rise to the aforementioned tensions. But it has been noted in the literature that the DESI DR2 BAO data is inconsistent with ACT CMB data at a $3.1\sigma$ level~\cite{SPT-3G:2025bzu, Ferreira:2025lrd}. Similar tensions exist between SPT and DESI. The tension between these two datasets is therefore larger than the tension between their combined value for $n_s$ and the typical predictions of Starobinsky's model, considering the na\"ive range $N_* = 50-60$. One should then be cautious against drawing strong conclusions about the status of the model before this BAO--CMB tension is resolved.

In any case, even taking these recent results at face value, one must recall that the value of $N_*$ (and hence of $n_s$) actually depends on the details of the reheating epoch, such as the EoS of the fluid and the final reheating temperature. As mentioned above, the typically cited range of $N_*=50-60$ considers only a vanilla scenario where the Universe behaves like matter (or at most like radiation) during reheating. 
But if the Universe was dominated by some kind of stiff matter, with an EoS parameter $w_\text{reh}>1/3$, then $N_*$ increases and hence the spectral index is pushed to slightly larger values than na\"ively thought, possibly bringing it back into agreement with the recent P-ACT-LB2-BK18 dataset~\cite{Drees:2025ngb, Zharov:2025zjg}. The problem is that, since in a stiff-dominated epoch the spectrum of inflationary gravitational waves becomes blue-tilted, a long stiff-dominated epoch (such as might be needed to sufficiently increase the prediction of $n_s$) could violate BBN bounds concerning the amount of radiation present during the formation of the first nuclei---and indeed a closer analysis has shown that the entire $1\sigma$ allowed region in the $({w}_\text{reh}, T_\text{reh})$ plane has already been excluded by this bound~\cite{Haque:2025uis}.

Now, the analyses mentioned above leave two gaps which we aim to fill in this paper. First, since the spectrum is blue-tilted, these primordial gravitational waves may enter the range of detectability of near-future interferometers, such as LISA and Einstein Telescope (ET), as well as of DECIGO and BBO. Still in the context of Starobinsky inflation, we investigate the range of reheating parameters that could be probed by these experiments. We find that, for a constant EoS parameter $w_\text{reh}$, most of parameter space accessible to LISA and ET is already excluded by BBN, and not much more information can be gained from these experiments. However, future experiments such as DECIGO and BBO could probe a significant portion of the region still allowed by BBN and agreeing with CMB+BAO at $2\sigma$ level---specifically they would be able to probe reheating temperatures up to $T_\text{reh}\sim 10^5-10^6$~GeV and $w_\text{reh}\gtrsim 0.35$.

Furthermore, we consider the scenario where the fluid during reheating is not constantly stiff, but starts as a matter-like fluid and later decays into a stiff kind of matter. This is the most realistic scenario in the Starobinsky model, since at the end of inflation the fluid energy density is known to behave like matter before it starts decaying into other particles. This is easy to see in the Einstein frame, where there appears an effective inflaton field which, at the end of inflation, oscillates around the bottom of an approximately quadratic potential, implying an EoS parameter $w=0$. But even in the Jordan frame it can be shown that, before the production of other particles start to dominate during reheating, the scale factor grows as in a matter-dominated Universe~\cite{Dorsch:2026ref}. It is therefore reasonable to assume that a more realistic reheating scenario with stiff matter involves first a matter-dominated period.

It could seem that this matter domination before a stiff epoch would be safer from BBN constraints, because, during matter domination, the GW spectrum is red-tilted, so the growth in GW energy density at high frequencies would be hampered. But our findings actually reveal the opposite: with the inclusion of a matter-dominated phase preceding the stiff phase, both the BBN constraints and the regions that can be probed by GW detectors become larger in the space of $(\overline{w}_\text{reh},T_\text{reh})$---where now $\overline{w}_\text{reh}$ is the \emph{average} EoS during reheating. This is because matter domination has $w=0$, so, in order to have a large average $\overline{w}_\text{reh}$, the stiff epoch must be closer to $w\approx 1$, leading to a sharper growth of the spectrum. So, also in this scenario, the $1\sigma$ concordance region with CMB+BAO is entirely excluded by BBN, while interferometers are able to probe a larger region of the remaining parameter space.

We also analyzed the case where reheating is first dominated by radiation, followed by a stiff fluid which later decays again into radiation to kickstart the $\Lambda$CDM history. Again we find that the $1\sigma$ region allowed by the ACT analysis is excluded by BBN, and we show the detectability of the remaining parameter space to GW detectors. Overall, our conclusion is that the $1\sigma$ region is excluded regardless of the details of reheating. The $2\sigma$ region might be excluded or not, and when not excluded by BBN, part of it could be probed by GW interferometers, rendering them important probes of the detailed dynamics of post-inflationary reheating.

This paper is organized as follows. Section \ref{sec:reh} reviews the theory of Starobinsky inflation and reheating and highlights the relevant parameters for our analysis. Section \ref{sec:stats} presents our setup and simulation results in extracting inflationary and reheating parameters from the P-ACT-LB2-BK18 dataset. Section \ref{sec:gws} overviews the properties of inflationary gravitational-wave signals, highlighting the impact of different EoS parameters during reheating on the tilting of the spectrum. Section \ref{sec:results} contains our main results, putting together the regions of the reheating parameter space favored by the CMB+BAO data, excluded by the BBN bound, and probeable by future gravitational-wave observatories. In section \ref{sec:conclusion} we share our conclusions.

\section{Starobinsky Inflation and Reheating}
\label{sec:reh}

The Starobinsky model of inflation is formulated in the so-called Jordan Frame through the action
\begin{equation}
    S_J = \frac{M_{\text{Pl}}^2}{16\pi} \int d^4x \sqrt{-g} \left( R + \frac{R^2}{6M^2} \right),
\end{equation}
where $M_{\text{Pl}}\equiv1/\sqrt{G}$ is the Planck Mass and $M$ is the mass scale of the quadratic term.

The complexity of solving the equations of motion in the Jordan Frame motivates us to perform a conformal transformation to the Einstein Frame, which casts the action into the simpler form
\begin{equation}
    S_E =  \int d^4x \sqrt{-g} \left[\frac{M_{\text{Pl}}^2}{16\pi}\bar{R}  - \frac{1}{2}(\nabla\phi)^2 - V(\phi)\right],
\end{equation}
with $\bar{R}$ the Ricci scalar in the Einstein frame and $\phi$, a field related to the conformal transformation (to be interpreted as the inflaton) with potential
\begin{equation}
    V(\phi) = V_0 \left( 1 - e^{-\sqrt{\frac{16\pi}{3}}\frac{\phi}{M_{\text{Pl}}}} \right)^2
    \label{eq:V}
\end{equation}
and $V_0 \equiv \frac{3}{32\pi}M^2M_{\text{Pl}}^2$ being the potential amplitude.

In the standard treatment, one solves the background and perturbation equations (more simply done via the slow-roll approximation) to find the inflationary observables. This methodology typically does not explicitly model the reheating epoch, which could alleviate the tensions with the updated P-ACT-BK-LB2 joint analysis. However, the inclusion of the post-inflationary epoch influences the total expansion history of primordial modes and directly affects the pivot $e$-fold number $N_*$.

To quantify these effects, we rewrite the Hubble crossing condition, $a_*H_*/k_* = 1$ as
\begin{equation}\label{eq:horizon_cross_mod}
    \frac{H_*}{k_*} \frac{a_*}{a_\text{end}}\frac{a_\text{end}}{a_\text{reh}}\frac{a_\text{reh}}{a_0}a_0 = 1,
\end{equation}
where the subscripts $*$, $\text{end}$, $\text{reh}$, and $0$ correspond to the quantities evaluated at the pivot $e$-fold number, the end of inflation, the end of reheating, and today, respectively. This reformulation decomposes the expansion history into a series of epochs for which we know the ratio $a_i/a_j$.

By definition, it follows that $a_*/a_\text{end} = e^{-N_*}$. The ratio $a_\text{reh}/a_0$ is found via the conservation of entropy $g_{*s,\text{reh}}a_\text{reh}^3T^3_\text{reh} = g_{*s,0}a_0^3T_0^3$, where $g_{*s,\text{reh}} = 106.75$ and $g_{*s,0} = 3.91$ are the effective number of relativistic degrees of freedom for entropy at the end of reheating\footnote{In the lower end of the range of reheating temperatures considered in our study $g_{*s,\text{reh}}<106.75$. Moreover, beyond-Standard-Model degrees of freedom could contribute at high temperatures. Nonetheless, since we have verified that our results depend weakly on this parameter, we set $g_{*s,\text{reh}}=106.75$ for all cases for simplicity.} and today, respectively, and $T$ is the temperature. Thus,
\begin{align}
    \frac{a_\text{reh}}{a_0} = \frac{T_0}{T_\text{reh}}\left(\frac{g_{*s,0}}{g_{*s,\text{reh}}}\right)^{1/3}.
    \label{eq:ratio1}
\end{align}

The ratio $a_\text{end}/a_\text{reh}$ is found via the continuity equation $\dot{\rho} + 3H\rho(1+w_\text{reh}) = 0$, where $w_\text{reh} \equiv P/\rho$ is the (possibly time-dependent) EoS parameter during reheating. Integrating this from the end of inflation to the end of reheating relates the energy densities to the scale factors,
\begin{equation}
    \frac{\rho_\text{reh}}{\rho_\text{end}} = \exp{\left[-\int_{a_\text{end}}^{a_\text{reh}} \frac{da}{a} \ 3(1+w_\text{reh})\right]}.
\end{equation}
Changing the time variable to the $e$-fold number and introducing the effective EoS
\begin{equation}
    \overline{w}_\text{reh} \equiv \frac{1}{N_\text{reh}-N_\text{end}}\int^{N_\text{reh}}_{N_\text{end}}w_\text{reh}(N)dN
    \label{wbar}
\end{equation}
we find
\begin{equation}\label{eq:ratio2}
    \frac{a_\text{end}}{a_\text{reh}} = \left(\frac{\rho_\text{reh}}{\rho_\text{end}}\right)^{\frac{1}{3(1+\overline{w}_\text{reh})}}.
\end{equation}

Because reheating marks the transition to a radiation-dominated Universe, the energy density at the end of this epoch is simply
\begin{equation}
    \rho_\text{reh} = \frac{\pi^2}{30}g_{*,\text{reh}}T_\text{reh}^4.
    \label{eq:energy_dens_reh}
\end{equation}
Substituting eqs. (\ref{eq:ratio1}) and (\ref{eq:ratio2}) and the definition $e^{-N_*}=a_*/a_{\text{end}}$ into eq. (\ref{eq:horizon_cross_mod}), we can solve for the pivot $e$-fold number $N_*$ to find
\begin{align}
    N_* = \ln\left[\frac{H_*a_0}{k_*}\left(\frac{\rho_\text{reh}}{\rho_\text{end}}\right)^{\frac{1}{3(1+\overline{w}_\text{reh})}}\frac{T_0}{T_\text{reh}}\left(\frac{g_{*s,0}}{g_{*s,\text{reh}}}\right)^{1/3}\right].
\end{align}
This equation can be further simplified by introducing the \textit{rescaled reheating parameter} \cite{Martin_2006, Martin_2014}
\begin{align}\label{eq:Rreh}
    R_{\text{reh}} &=\frac{a_{\text{end}}}{a_{\text{reh}}} \left( \frac{\rho_{\text{end}}}{\rho_{\text{reh}}} \right)^{\frac{1}{4}} \frac{\rho_{\text{end}}^{\frac{1}{4}}}{M_{\text{Pl}}}\notag\\
    &= \left( \frac{\pi^2}{30} g_{*,\text{reh}} \frac{T_{\text{reh}}^4}{\rho_{\text{end}}} \right)^{\frac{1}{3(1+\overline{w}_{\text{reh}})}-\frac{1}{4}} \frac{\rho_{\text{end}}^{\frac{1}{4}}}{M_{\text{Pl}}},
\end{align}
which contains all the macroscopic effects of reheating. Substituting eq.~(\ref{eq:energy_dens_reh}), $a_0 = 1$, $k_* = 0.05$ Mpc$^{-1}$, $T_0 = 2.7255$~K \cite{ACTDR6, Fixsen_2009}, and $\rho_\text{end} = 3 M_\text{Pl}^2 H^2_\text{end}/8\pi$, we finally obtain
\begin{equation}
    N_* = 62.7 + \ln{R_\text{reh}} + \ln\left(\frac{H(N_*)}{H_\text{end}}\right).
    \label{eq:Nstar_raw}
\end{equation}
Given $\ln V_0$ one can solve the background equations numerically to determine $H_\text{end}$ and $\rho_\text{end}$. Then, supplemented with a given $\ln R_\text{reh}$, this equation can be solved numerically for $N_*$ using fixed-point iteration methods.

Since the Starobinsky scenario is an instance of a slow-roll model of inflation, the slow-roll parameters can be easily calculated from the potential in eq.~\eqref{eq:V} and the spectral index can be expressed as~\cite{DeFelice:2010aj}
\begin{equation}
    n_s \approx 1-\dfrac{2}{N_*}.
\end{equation}
For $N_*=62.7$ one finds $n_s\approx 0.968$, which is in a $\sim 2\sigma$ tension with the ACT DR6 measurement of $n_s=0.9752 \pm 0.0030$. A better agreement would require a larger $N_*$, i.e. more inflationary $e$-folds. In order to understand how this requirement translates into a constraint on the reheating dynamics, let us rewrite eq.~\eqref{eq:Nstar_raw} as
\begin{equation}
    \small
    N_* = 62.7 + \frac{1-3\overline{w}_\text{reh}}{12(1+\overline{w}_\text{reh})}\ln\frac{\rho_\text{reh}}{\rho_\text{end}} + \ln \frac{\rho_\text{end}^{\frac{1}{4}}}{M_\text{Pl}} + \ln \frac{H(N_*)}{H_\text{end}}.
    \label{eq:Nstar}
\end{equation}

The third term in eq.~\eqref{eq:Nstar} is negative because the scale of inflation in Starobinsky's model is sub-Planckian. On the other hand, the last term  provides a positive but small contribution to $N_*$. This is because during Starobinsky inflation the Hubble parameter remains \emph{approximately} constant, while actually strictly decreasing\footnote{Recall that, during the slow-roll regime, $H^2 = 8\pi V(\phi)/3M_\text{Pl}^2$. During inflation, the potential is approximately flat, so $H\approx \text{constant}$, but the field rolls \emph{down} the potential, so $H$ strictly decreases.}, so its value $H(N_*)$ when the pivot mode $k_*$ crosses the Hubble radius is larger than $H_\text{end}$ at the end of inflation. 

We thus see that, in order to push $n_s$ towards larger values, compatible with ACT DR6 data, the second term must provide a significant positive contribution to $N_*$. Since $\rho_\text{reh} \leq \rho_\text{end}$, this means we must have $\overline{w}_\text{reh}>1/3$, and therefore an epoch of reheating dominated (on average) by a stiff fluid.

\section{Statistical Analysis of Starobinsky reheating in light of ACT DR6}
\label{sec:stats}
In this section we make this discussion more quantitative by showing the allowed regions in the space of reheating parameters in light of ACT DR6 data. To do this, we perform a MCMC sampling using the \texttt{Cobaya} framework~\cite{Torrado_2021, 2019ascl_soft10019T} and a modified version of the Boltzmann code \texttt{CLASS} ~\cite{Diego_Blas_2011}. The input configuration is listed in Appendix A. The background equations and the primordial power spectrum are solved numerically without slow-roll approximations, and a bissection method is also implemented to find $N_*$ via eq.~(\ref{eq:Nstar_raw}).

To study the Starobinsky model, we use the P-ACT-LB2-BK18 dataset, with corresponding likelihoods detailed in Table~\ref{tab:likelihoods}. To ensure proper convergence, we ran 8 chains with a Gelman-Rubin $R-1$ convergence criterion \cite{GRRminus1} of $0.01$ on the means and $0.05$ on the $2\sigma$ confidence limits.

\begin{table}
    \centering
    \begin{tabular}{ll}
        \hline
        \multicolumn{1}{c}{Likelihood} & \multicolumn{1}{c}{Data} \\
        \hline
        \texttt{planck\_2018\_lowl.}    & Low-$\ell$ E-mode \\
        \quad \texttt{EE\_sroll2}   & \quad polarization \\
        \texttt{planck\_2018\_lowl.TT}  & Low-$\ell$ temperature \\
        \texttt{act\_dr6\_cmbonly.} & Foreground-cleaned cut \\
        \quad \texttt{PlanckActCut} & \quad Planck high-$\ell$ CMB \\
        \texttt{act\_dr6\_cmbonly}  & Foreground-cleaned \\
                                & \quad high-$\ell$ CMB\\
        \texttt{act\_dr6\_lenslike} & Planck and ACT lensing \\
        \texttt{bao.desi\_dr2.} & DESI DR2 BAO \\
        \quad \texttt{desi\_bao\_all}   & \\
        \texttt{bicep\_keck\_2018}  & CMB B-mode \\
        \hline
    \end{tabular}
    \caption{Likelihoods used in the MCMC run.}
    \label{tab:likelihoods}
\end{table}

\begin{figure}
    \centering
    \includegraphics[width=1.0\linewidth]{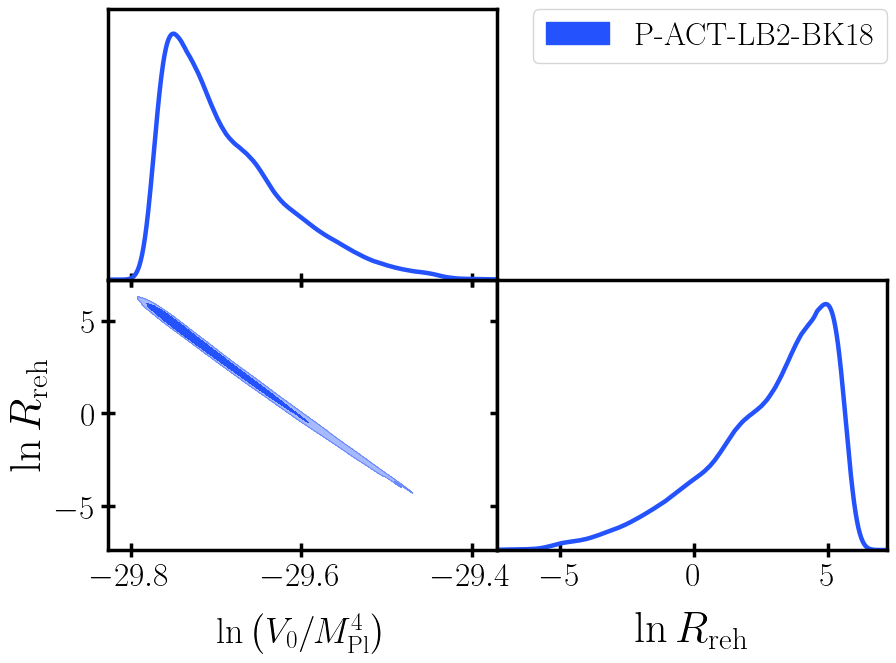}
    \caption{Triangle plot of the $\ln R_\text{reh} \times \ln V_0$ plane and their marginalized posteriors.}
    \label{fig:lnRxlnV0}
\end{figure}

Following the computational pipeline presented in \cite{km3q-rm34}, we initially sample the parameter plane $\theta = (\ln{V_0}, \ln{R_\text{reh}})$ assuming flat priors within the ranges $\ln{V_0} \in [-150, -28]$ and $\ln{R_\text{reh}} \in [-48.5805, 16.1935 + (1/3)\rho_{\text{end}}]$. The limits on $\ln{R_\text{reh}}$ are derived by imposing the constraints $T_\text{reh} \geq 4$~MeV~\cite{PhysRevD.70.043506} and $-1/3 < \overline{w}_\text{reh} < 1$, where the lower bound on $\overline{w}_\text{reh}$ ensures that inflation has ended and the upper bound is imposed to avoid superluminal adiabatic perturbations. The resulting plot is presented in fig. \ref{fig:lnRxlnV0}, and the best-fit and mean values with $1\sigma$ and $2\sigma$ uncertainties are shown in Table \ref{tab:means}.

\begin{table}
    \centering
    \vspace{0.1cm}
    \renewcommand{\arraystretch}{1.3}
    \begin{tabular}{l @{\hspace{0.7cm}} l @{\hspace{0.7cm}} l}
        \hline
        Parameter & Best-fit & Mean $\pm$ $1\sigma$ \\
        \hline
        $\ln R_{\text{reh}}$    & $5.587$    & $2.8^{+2.9}_{-1.1}$ \\
        $\ln \frac{V_0}{M_{\text{Pl}}^4}$   & $-29.771$  & $-29.688^{+0.031}_{-0.086}$    \\[1.5mm]
        \hline
        $n_s$   & $0.9719$  & $0.9706^{+0.0013}_{-0.00043}$  \\
        $\ln(10^{10}A_s)$   & $3.062$   & $3.0634 \pm 0.0042$  \\
        $10^3 r$    & $2.300$   & $2.502^{+0.070}_{-0.21}$  \\
        $N_*$   & $69.0$    & $66.2^{+2.9}_{-1.1}$ \\
        $H(N_*)$[GeV] & $1.200 \cdot 10^{13}$ & $\left(\,1.251^{+0.018}_{-0.054}\,\right)\cdot 10^{13}$\\
        $H_\text{end}$[GeV] & $5.858\cdot 10^{12}$ & $\left(\,6.111^{+0.090}_{-0.269}\,\right)\cdot 10^{12}$\\
        $\ln \frac{H(N_*)}{H_\text{end}}$   & $0.71684$    & $0.71647^{+0.00046}_{-0.00016}$  \\[1.5mm]
        \hline
        $\chi^2_{\text{total}}$ & $1383.90$& \dotfill \\
        $\chi^2_{\text{Planck-lowEE}}$   & $391.63$ & \dotfill \\
        $\chi^2_{\text{Planck-lowTT}}$   & $22.37$ & \dotfill \\
        $\chi^2_{\text{ACT-CMB}}$   & $160.44$  & \dotfill \\
        $\chi^2_{\text{ACT-CMB.PlanckCut}}$   & $220.70$  & \dotfill \\
        $\chi^2_{\text{ACTlens}}$   & $19.74$   & \dotfill \\
        $\chi^2_{\text{DESI-BAO}}$  & $11.74$   & \dotfill \\
        $\chi^2_{\text{BK18}}$  & $536.71$  & \dotfill \\
        \hline
    \end{tabular}
    \caption{Best-fit (rounded) and mean values with 68\% confidence level (CL) for the primary and derived cosmological parameters, and $\chi^2$ values for the used likelihoods. We used the BOBYQA \cite{10.1145/3338517} algorithm to maximize the posterior and find the best-fit values.}
    \label{tab:means}
\end{table}

After obtaining the chains, we use the kernel density estimator (kde) of the \texttt{getdist} package \cite{Lewis_2025} with a smoothing factor of 0.2 to produce the joint posterior $P(\ln{V_0}, \ln{R_\text{reh}})$ and the marginalized pdfs of each parameter. The results, shown in fig.~\ref{fig:lnRxlnV0}, are in excellent agreement with those in ref.~\cite{km3q-rm34}.

\begin{figure}
    \centering
    \includegraphics[width=1.\linewidth]{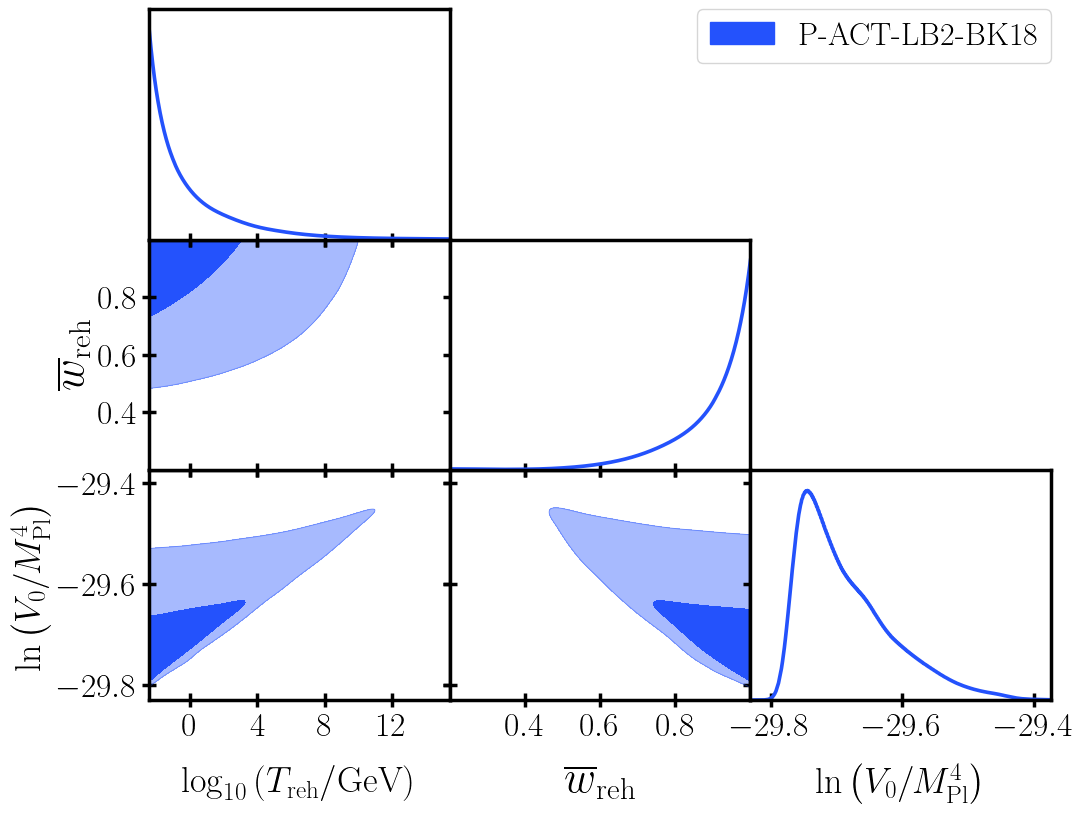}
    \caption{Triangle plot of the inflationary potential scale $\ln{\left(V_0/M_{pl}^4\right)}$, the reheating temperature converted to a logarithmic scale of base 10, $\log_{10}\left({T_\text{reh}}/\text{GeV}\right)$, and EoS $\overline{w}_\text{reh}$.}
    \label{fig:triangle}
\end{figure}

In the second part of the analysis, the posterior $P(\ln{V_0}, \ln{R_\text{reh}})$ is used as the likelihood for a new MCMC on the 3D parameter space $\theta' = (\ln{V_0}, \ln{T_\text{reh}},\overline{w}_\text{reh})$, with flat priors in the ranges $\ln T_\text{reh}/M_{pl} \in [-49.4702, -8.5]$ and $\overline{w}_\text{reh} \in [-1/3,1]$. The upper bound on $T_\text{reh}$ is found by imposing $\rho_\text{reh} \leq \rho_\text{end}$. Our MCMC routine slices the 3D parameter space into 200 $\ln{T_\text{reh}}\times \overline{w}_\text{reh}$ planes by proposing steps in the $\ln{V_0}$ axis, which are weighted based on the marginalized posterior of $\ln{V_0}$ obtained in the first phase. This ensures that the probability distribution of $\ln{V_0}$ is preserved in the second scan. Finally, each plane is explored using the python library \texttt{emcee} \cite{Foreman-Mackey_2013}, configured to use 32 walkers, 10,000 points per slice and a 30\% burn-in. The resulting triangle plot is shown in fig.~\ref{fig:triangle}, and the corresponding 68\% and 95\% CL constraints are summarized in Table~\ref{tab:means_fundamental_params}.

\begin{table}
    \renewcommand{\arraystretch}{1.5}
    \centering
    \vspace{0.1cm}
    \begin{tabular}{c @{\hspace{0.7cm}} l @{\hspace{0.7cm}} l}
       \hline
        Parameter & $1\sigma$ Constraints & $2\sigma$ Constraints \\
        \hline
        $\overline{w}_{\text{reh}}$  & $> 0.864$  & $> 0.599$  \\
        $\ln \left(V_0/M_{\text{Pl}}^4\right)$ & $-29.686^{+0.031}_{-0.086}$  & $-29.686^{+0.15}_{-0.10}$  \\
        $\ln \left(T_{\text{reh}}/M_\text{Pl}\right)$  & $< -42.9$  & $< -29.8$  \\[1mm]
        \hline
        $\log_{10}\left(T_\text{reh}/\text{GeV}\right)$ & $< 0.451$  & $< 6.14$  \\
        \hline
    \end{tabular}
    \caption{Constraints on the fundamental inflation and reheating parameters. We report the mean and $1-2\sigma$ uncertainties for $\ln \left(V_0/M_{\text{Pl}}^4\right)$, while providing the $68\%$ and $95\%$ CL unilateral limits for the effective EoS, $\overline{w}_\text{reh}$, and reheating temperature, $\ln{(T_\text{reh}/M_\text{pl})}$ and $\log_{10}\left(T_\text{reh}/\text{GeV}\right)$.}
    \label{tab:means_fundamental_params}
\end{table}

The results obtained complement well the ones from refs.~\cite{km3q-rm34, Haque:2025uis}, which used the P-ACT-LB-BK18 mean values and likelihoods. As expected, the P-ACT-LB2-BK18 dataset constrains parameter space more tightly since it favors a higher spectral index ($\bar{n}_s = 0.9752$) when compared to P-ACT-LB-BK18 ($\bar{n}_s = 0.9743$), and so requires values of $\ln V_0$ and $\ln R_\text{reh}$ that yield a larger number of $e$-folds $N_*$. Thus, given the direct correlation between $\ln R_\text{reh}$ and $N_*$, the allowed combinations of the inflationary energy scale and the reheating thermal history become more restricted, decreasing the probability density at the tails.

We see that Starobinsky inflation could only agree with the ACT DR6 + DESI DR2 data if the average EoS during reheating is indeed very stiff and the reheating temperature is not too large. For the marginalized posterior, we find $\overline{w}_\text{reh} \gtrsim 0.86$ and $T_\text{reh}\lesssim 3$~GeV at $1\sigma$  ($\overline{w}_\text{reh}\gtrsim 0.6$ and $T_\text{reh}\lesssim 10^{6}$~GeV at $2\sigma$). Furthermore, the pivot value $N_*$ is limited to the interval $[55.2,70.4]$ at the $5 \sigma$ CL, which increases the allowed region of Hubble radius exit.

\section{Primordial Gravitational Waves}
\label{sec:gws}
Primordial gravitational waves are a crucial prediction of inflationary models, and their detection could provide invaluable information about the early Universe \cite{Maggiore:2018sht,Maggiore:1999vm,Guzzetti:2016mkm,alves2026gravitationalwavesbigbang}. In a standard single-field slow-roll inflationary scenario, such as ours, this stochastic background is generated by tensor perturbations of the metric, arising from quantum fluctuations of the gravitational field.

The evolution of primordial tensor perturbations $h_k\equiv h_+, h_\times$ is governed by
\begin{equation}
h_k''+2\mathcal H h_k'+k^2 h_k=0,
\end{equation}
where primes denote derivatives with respect to conformal time $\tau$, and $\mathcal H \equiv a'/a$. Defining the canonical variable
\begin{equation}
    \mathfrak{h}_k \equiv a h_k,
\end{equation}
the tensor equation can be rewritten as
\begin{equation}
    \mathfrak{h}_k''+\left(k^2-\frac{a''}{a}\right)\mathfrak{h}_k=0.
\end{equation}

For super-Hubble modes, $k\ll aH$, the tensor amplitude remains approximately constant. After Hubble radius reentry, $k\gg aH$, the term $a''/a$ becomes negligible and the solution reduces to $\mathfrak{h}_k \simeq e^{\pm ik\tau}$,
which implies
\begin{equation}
    |h_k| \propto a^{-1} \quad \text{(sub-Hubble modes).}
    \label{eq:modedecay}
\end{equation}

The quantum fluctuations of these tensor modes are commonly described by their variance, in terms of which the dimensionless power spectrum is defined:
\begin{equation}
    \left\langle(h_{ij}^{\mathrm{TT}})^\dagger (\tau,\vec x)h^{ij}_{\mathrm{TT}}(\tau,\vec x)\right\rangle\equiv\int d(\ln k) \Delta^2_h(k,\tau).
\end{equation}
In slow-roll inflation, $\Delta^2_h\propto k^{n_T}$, with a tensorial spectral index $n_T=-2\epsilon=-r/8$, where $\epsilon$ is a slow-roll parameter \cite{alves2026gravitationalwavesbigbang}. Therefore, the inflationary spectrum of gravitational waves is originally sourced nearly flat, and we henceforth set $n_T=0$. However, as we shall discuss next, the spectrum measurable in the present picks up a frequency dependence determined by the epoch in which different modes re-entered the Hubble sphere.

The energy density carried by gravitational waves is given by~\cite{Misner:1973prb}
\begin{align}
    \rho_{\text{gw}} &= \frac{1}{32\pi G}\left\langle\partial_t h_{ij}^{\text{TT}}\partial_t h^{ij}_{\text{TT}}\right\rangle,
\end{align}
where the label $\mathrm{TT}$ indicates the transverse-traceless gauge. The wavenumber-dependent density parameter of gravitational waves is then \cite{alves2026gravitationalwavesbigbang}
\begin{equation}
    \Omega_\text{gw}(k)\equiv\frac{1}{\rho_c}\frac{d\rho_\text{gw}}{d\ln k}=\frac{k^2\Delta_{h,\text{prim}}^2(k)}{12 a(\tau)^2H(\tau)^2}T_h^2(k,\tau),
    \label{eq:Ogwspectrum}
\end{equation}
where $\rho_c=3H^2/8\pi G$ is the critical density, $\Delta^2_{h,\text{prim}}(k)$ is the primordial dimensionless power spectrum, and the transfer function $T_h(k,\tau)$ encapsulates the time evolution of each mode---which only starts at Hubble radius reentry, given that the amplitude of super-Hubble modes does not evolve with time.

The transfer function is $k$-dependent through a factor $T_h^2(k,\tau)\propto \left(a(\tau_k)/a(\tau)\right)^2$, where $\tau_k$ is the Hubble radius reentry time of mode $k$, so that $a(\tau_k) = k/H(\tau_k)$ \cite{duval2024investigatingcosmichistoriesstiff}. When the Universe is dominated by a component whose EoS parameter is $w$, the Friedmann equation informs that $a(t)\propto t^{2/3(1+w)}$ so that $ k=aH\propto t^{-(1+3w)/3(1+w)}$. Therefore, neglecting the small scale-dependence of the primordial spectrum $\Delta_{h,\text{prim}}(k)$, the $k$-dependence of $\Omega_\text{gw}$ in eq.~\eqref{eq:Ogwspectrum} as measured in the present is 
\begin{equation}
\Omega_\text{gw}(k)
\propto
a(k)^2k^2
\propto
k^{\frac{2(3w-1)}{3w+1}}.
\label{eq:GWscaling}
\end{equation}

In the expression above, $w$ is the value of the EoS of the dominant fluid when the mode with frequency $f$ re-enters the Hubble sphere. Hence, the present-day stochastic gravitational-wave background acquires a characteristic tilt determined by the effective EoS during reheating. The full expression, taking the proportionality terms into account, can be reconstructed by noticing that at the end of reheating the Universe is radiation-dominated and the spectrum is flat, with amplitude given by~\cite{Boyle:2007zx, Haque:2021dha}
\begin{equation}
    \Omega_\text{gw}^\text{(rd)} h^2
    = 
    \dfrac{rA_s\Omega_{\text{rad},0}h^2 }{24} 
    \left(\dfrac{g_{*,\text{reh}}}{g_{*,0}}\right)
    \left(\dfrac{g_{*s,0}}{g_{*s,\text{reh}}}\right)^{4/3}
\end{equation}
with $A_s = 2.12\times 10^{-9}$~\cite{ACTDR6_cosmo_param}, $\Omega_{\text{rad},0}h^2 = 2.47\times 10^{-5}$, $g_{*,0}=3.36$, $g_{*,\text{reh}} = 106.75$ and $r=12/N_*^2$~\cite{DeFelice:2010aj}. In particular, notice that the value of $r$ also depends on $\overline{w}_\text{reh}$ through the dependence in $N_*$ (cf. eq.~\eqref{eq:Nstar}).

\begin{figure}
    \centering
    \includegraphics[width=0.45\textwidth]{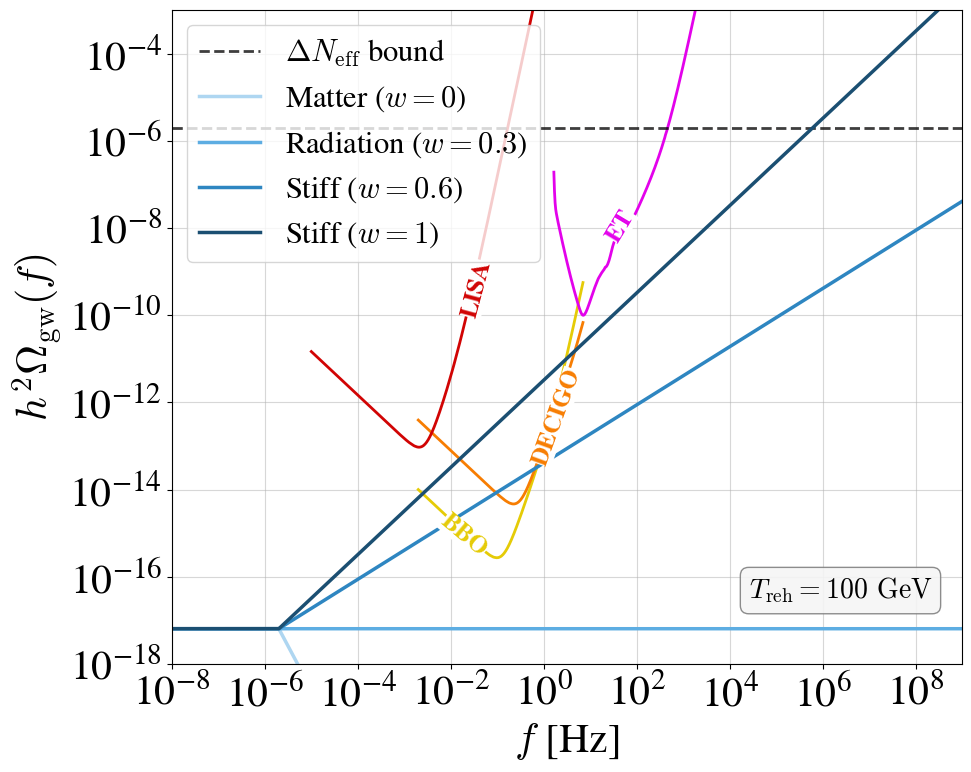}
    \caption{Gravitational-wave spectra (blue lines) for various constant EoS parameters during reheating, keeping the reheating temperature fixed at $T_\text{reh}=100~\text{GeV}$. The effect of increasing (resp. decreasing) the reheating temperature is to shift to the right (resp. left) the point where the spectrum transitions from flat to tilted. The curves in red (LISA), orange (DECIGO), yellow (BBO), and pink (Einstein Telescope--ET) represent the sensitivity curves for these gravitational-wave experiments. The curves are shown here for illustrative purposes only, as the rigorous detectability analysis will be conducted by computing the signal-to-noise ratio. }
    \label{fig:GW spectrum to only sd}
\end{figure}

One sees that, for $w>1/3$ (stiff matter), the spectrum is blue-tilted (grows with frequency), whereas for radiation domination ($w=1/3$) the spectrum is flat, recovering the expected scale-invariant behaviour. This is illustrated in fig.~\ref{fig:GW spectrum to only sd}. The frequency where the spectrum changes from flat to blue-tilted is characterized by the reheating temperature, given explicitly by~\cite{Nakayama:2008wy}
\begin{equation}
    f_\text{reh} = 0.026~\text{Hz}\left(\frac{g_{*,\text{reh}}}{106.75}\right)^{1/6} \left(\frac{T_\text{reh}}{10^6~\text{GeV}}\right).
    \label{eq:freh}
\end{equation}
The spectrum goes up to a maximum frequency $f_\text{end}$ corresponding to a mode $k_\text{end}=a_\text{end}H_\text{end}$ that crosses the Hubble radius at the end of inflation\footnote{Modes above this frequency never leave the Hubble sphere and never get frozen out, so they get diluted away by inflation, always decaying according to eq.~\eqref{eq:modedecay}.}. After redshifting to today, and repeating some steps performed in section~\ref{sec:reh}, this frequency $f_\text{end}=a_\text{end} H_\text{end}/(2\pi a_0)$ becomes
\begin{equation}\begin{split}
    f_\text{end} 
    &= \dfrac{H_\text{end}}{2\pi}
        \left(\frac{\rho_\text{reh}}{\rho_\text{end}}\right)^{\frac{1}{3(1+\overline{w}_\text{reh})}}
        \frac{T_0}{T_\text{reh}}\left(\frac{g_{*s,0}}{g_{*s,\text{reh}}}\right)^{1/3}.
    \label{eq:fend}
\end{split}\end{equation}
The values of $H_\text{end}$ and $\rho_\text{end}$ are obtained numerically from solving the background equations once $V_0$ is fixed, whereas $\rho_\text{reh}$ is given in terms of $T_\text{reh}$ by eq.~\eqref{eq:energy_dens_reh}. We thus reach the important result that, in this case, the spectrum is totally fixed by knowledge of $V_0$, $T_\text{reh}$, and $\overline{w}_\text{reh}$.

In fig.~\ref{fig:GW spectrum to only sd}, we illustrate the impact of the EoS on the gravitational-wave spectrum. Introducing a constant EoS $\frac{1}{3} < w < 1$ during reheating, which is associated with a stiff epoch, induces a blue tilt in the spectrum. This means that the spectrum's amplitude increases at higher frequencies, thereby enhancing its detectability. The experimental sensitivity curves in the figure were plotted using the data from refs. \cite{gwplotter_web,Moore_2015} for illustrative purposes only, as the rigorous detectability analysis will be conducted by calculating the signal-to-noise-ratio, as prescribed in the next section.

Primordial gravitational waves in a standard slow-roll inflationary model are nearly scale-invariant, as shown in references \cite{Guzzetti:2016mkm} and \cite{Nakayama:2008wy}. Therefore, the detection of a behavior like the one shown in fig. \ref{fig:GW spectrum to only sd} would be a strong signature of a phenomenology like the one explored in this work.  

\subsection{Signal-to-Noise-Ratio (SNR)}

A primary motivation for studying primordial gravitational waves is the possibility of their direct detection. The detectability of a model can be assessed using the signal-to-noise ratio (SNR). For a continuous spectrum, like the ones considered here (see e.g. fig.~\ref{fig:GW spectrum to only sd}) the SNR is computed as an integral of the energy density generated by the model, $\Omega_{\text{signal}}(f)$, over the sensitivity of the experiment, quantified by its noise density spectrum, $\Omega_{\text{noise}}(f)$. This accounts for cumulative effects of the signal over a frequency range: even if the amplitude itself at each frequency would always be below the experiment sensitivity, summing the ratio over many frequencies could lead to a sufficiently large SNR~\cite{Thrane:2013oya}. Also taking into account the observation time of the experiment $t_\text{obs}$, we finally have a signal-to-noise ratio 
\begin{equation}
    \text{SNR}=\left[ t_{\text{obs}} \int_{f_0}^{f_\text{end}} df \left( \frac{\Omega_{\text{signal}}(f)}{\Omega_{\text{noise}}(f)} \right)^{2} \right]^{1/2}
\end{equation}
with $f_0 = a_0 H_0/(2\pi)$ the frequency of the mode crossing the Hubble radius today (being therefore the mode with lowest detectable frequency).

To consider a detection we adopt the criterion that $\text{SNR}>10$.

\subsection{Observational Constraint: Big Bang nucleosynthesis}

Big Bang Nucleosynthesis is the primordial production of light elements through nuclear reactions in the hot medium that filled the radiation-dominated Universe. Its ability to explain current element abundances renders it a high-precision probe of early-Universe physics. 

To compute the abundance of light elements formed during this primordial nucleosynthesis, one needs to know the ratio of proton and neutron abundances at the onset of the process. While neutrons and protons are in chemical equilibrium with the plasma, their number densities follow a Boltzmann distribution $\propto e^{-E/T}$. Eventually, however, the Hubble expansion dominates over the weak-interaction rate $\Gamma_{pe\leftrightarrow n\nu} \sim G_F^2 T^5$,  and the ratio of number densities essentially freezes---apart from small changes coming from neutron decay---into $n_n/n_p\approx\exp{[(m_p-m_n)/T_f]}$, where $m_n$ and $m_p$ are the neutron and proton masses, and $T_f$ is the freeze-out temperature. Since the Hubble rate depends on the radiation energy density in the early Universe as $H\sim \sqrt{\rho_\text{rad}} \sim (g_* T^4)^{1/2}$, one sees that decoupling happens earlier when more relativistic degrees of freedom are present, leading to more neutrons being available to form heavier elements. This, in turn, affects the very well-measured abundances of light elements~\cite{Mossa:2020gjc,Hsyu:2020uqb,Cooke:2017cwo,Cyburt:2015mya}. Indeed, it is easy to see from the above argument that the freeze-out temperature should scale with the effective number of relativistic degrees of freedom as $T_f\sim g_*^{1/6}$\cite{Maggiore:1999vm,Kolb:1990vq}. 

At typical BBN temperatures of $\sim 1$ MeV, the relativistic Standard Model (SM) species are the photon, the electron, the positron, 3 neutrinos, and 3 antineutrinos (electronic, muonic and tauonic). Therefore, the effective number of relativistic degrees of freedom is $g_{*,\text{SM}}=2+(7/8)(2+2+3+3)=43/4$. On top of these SM degrees of freedom, there may be other contributions to the radiation energy density in this epoch, such as those from primordial GWs, which will effectively translate into additional contributions to this number of relativistic species in the plasma (since $\rho_\text{rad}\sim g_* T^4$). In fact one typically parametrizes this effect by defining $g_{*}=2+(7/8)(4+2N_\text{eff})$, where $N_\text{eff}\equiv 3+\Delta N_\text{eff}$ is called the \emph{effective number of neutrinos} in the early Universe. The naming stems from early attempts of using cosmological measurements to constrain the number of light neutrino species---an important and interesting interplay between cosmology and particle physics. However, $\Delta N_\text{eff}$ does not necessarily stem from neutrino physics: in fact, as we argue here, it can come from contributions of GWs to the radiation energy density at BBN. Observations of Helium-4 and Deuterium abundances reveal that $\Delta N_\text{eff}\lesssim0.4$, which is the bound we will use henceforth~\cite{Mossa:2020gjc,Hsyu:2020uqb,Cooke:2017cwo,Cyburt:2015mya}.

Assuming that, besides Standard Model particles, primordial GWs are the only additional relativistic species during BBN, one finds a bound on the integrated spectrum\footnote{Despite being a constraint on the integral of $\Omega_\text{gw}(f)$, in the literature eq.~\ref{BBNconstraint} is most often used as a bound on the value of $\Omega_\text{gw}(f)$ itself. These are indeed roughly equivalent for sufficiently smooth spectra, such as the ones considered here, where the dominant contribution to the integral over one unit of $\Delta\log f\sim 1$ typically comes from the neighborhood of the peak. Here we have checked that both approaches---considering the full integral or a bound on the peak value---yield identical results.} expressed as~\cite{Maggiore:1999vm}, 
\begin{equation}
    \int_{f_\text{BBN}}^{f_\text{end}}\mathrm d(\ln f)h^2\Omega_{\text{gw}}(f)\leq 5.6\times10^{-6}\Delta N_\text{eff},
    \label{BBNconstraint}
\end{equation}
where $H_0=100h$ $\mathrm{km\ s^{-1}\ Mpc^{-1}}$, $f_\text{BBN}\approx1.8\times10^{-11}$~Hz is the frequency of the tensor mode that crossed the Hubble radius during BBN~\cite{Boyle:2007zx}, and $f_\text{end}$ is given in eq.~\eqref{eq:fend}.

\subsection{A broken-power-law spectrum}
Looking at Fig.~\ref{fig:GW spectrum to only sd} one sees that a purely stiff-dominated reheating, with a constant EoS $w>1/3$, will lead to a violation of $\Delta N_\text{eff}$ bounds if $T_\text{reh}$ is sufficiently low and/or $w_\text{reh}$ is sufficiently large. For instance, one grasps from the figure that for $T_\text{reh}=100$~GeV and $w_\text{reh}=1$ the peak is above $\Omega_\text{gw}h^2 =10^{-4}$ and surely the integral~\eqref{BBNconstraint} will violate the upper bound set by $\Delta N_\text{eff}\lesssim 0.4$. For $w_\text{reh}=0.6$ the spectrum's peak is already close to $\Omega_\text{gw}h^2 \sim 10^{-7}$, thus nearly violating the BBN bound described above---a slightly larger $w_\text{reh}$ or slightly lower $T_\text{reh}$ would violate the bound, as will be confirmed in the results section below. In this case this happens because a purely stiff-dominated reheating leads to a constantly blue-tilted spectrum, steadily growing up to the region of $f\sim f_\text{end}$.

But this scenario assumes that, after the end of inflation, the inflaton energy density decays instantaneously into this stiff matter---which is not the most realistic modeling of reheating. In fact, in Starobinsky inflation it is known that the end of inflation is dominated by a matter-like fluid~\cite{Dorsch:2026ref}, as expected for an inflaton field oscillating around the bottom of an approximately quadratic potential. Then, during these oscillations, the inflaton decays into other forms of matter (possibly stiff-like) that eventually dominate the Universe. 

\begin{figure}
    \includegraphics[width=0.45\textwidth]{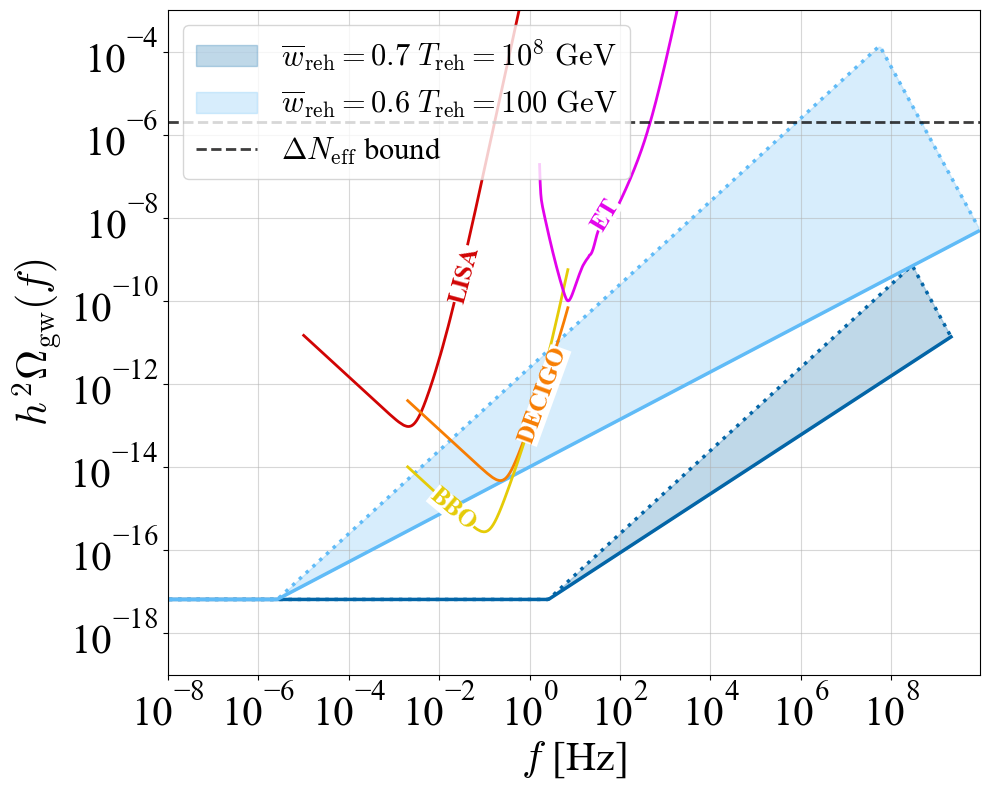}
    \caption{Similar to fig.~\ref{fig:GW spectrum to only sd}, but now with a broken power-law spectrum, corresponding to a two-stage reheating, with early matter domination followed by stiff fluid domination. Unlike the previous case, the spectrum here depends on an additional parameter $w_\text{sd}$---the EoS of the stiff epoch, which is now different from the average EoS $\overline{w}_\text{reh}$ that is actually constrained by ACT. We then have a continuum of spectral shapes within the range $\overline{w}_\text{reh} \leq w_\text{sd} \leq 1$. The solid lines represent the cases where $w_\text{sd} = \overline{w}_\text{reh}$, the dotted lines denote the other extreme case where $w_\text{sd} = 1$, and the shaded region between them encompasses the other possible scenarios. Since $w_\text{sd}\geq \overline{w}_\text{reh}$, the curves here are always significantly steeper than the corresponding ones in fig.~\ref{fig:GW spectrum to only sd}. This explains why the experimental bounds in this scenario exclude a larger portion of parameter space.}
    \label{fig:broken}
\end{figure}

\begin{figure}
    \includegraphics[width=0.45\textwidth]{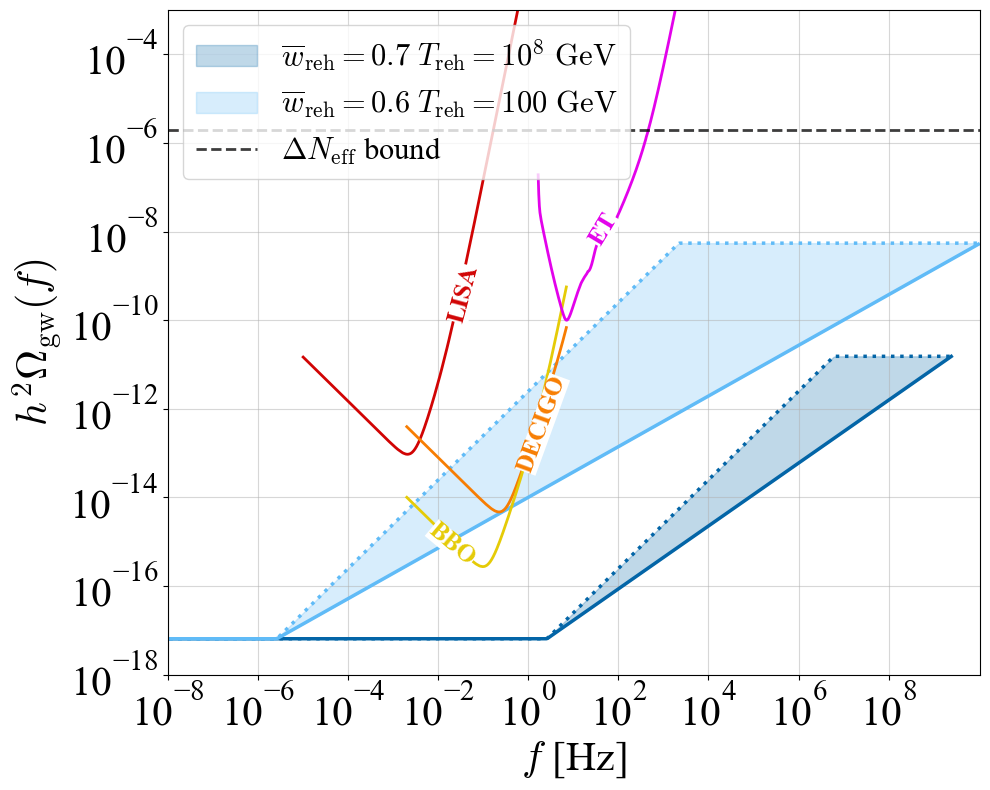}
    \caption{Same as fig.~\ref{fig:broken}, but for an early radiation-dominated era followed by a stiff-dominated epoch.
    }    \label{fig:rad}
\end{figure}

Considering the possibility of additional eras during reheating, we investigate two alternative cosmological histories. The first follows the sequence: inflation $\rightarrow$ matter domination $\rightarrow$ stiff domination $\rightarrow$ $\Lambda$CDM, with the resulting GW spectrum illustrated in fig.~\ref{fig:broken}. The second scenario is analogous but features an early radiation-dominated epoch instead of a matter-dominated one, as depicted in fig.~\ref{fig:rad}. Note that ACT can only place bounds on the average EoS parameter $\overline{w}_\text{reh}$ (cf. eq.~\eqref{eq:Nstar}), whereas the detailed history of reheating will impact the spectrum and hence the $\Delta N_\text{eff}$ bound and gravitational-wave detectability. In fact, one might na\"ively expect that adding a matter- or a radiation-dominated epoch could alleviate the $\Delta N_\text{eff}$ constraint, since the spectrum becomes red-tilted or scale-invariant before crossing the upper bound. However, we will show that this is not the case in section \ref{sec:results}.

These broken cases are characterized by parameters $f_\text{end}$ (related to $V_0$, $\overline{w}_\text{reh}$, and $T_\text{reh}$ via eq.~\eqref{eq:fend}), $f_\text{reh}$ (directly related to $T_\text{reh}$ via eq.~\eqref{eq:freh}), $\overline{w}_\text{reh}$, and one additional parameter, namely the frequency $f_\text{sd}$ where the Universe transitions from the first epoch (matter or radiation-dominated) to the stiff era. This frequency is directly linked to the EoS during the stiff epoch, $w_\text{sd}$, which varies within the range $\overline{w}_\text{reh} \leq w_\text{sd} \leq 1$. Consequently, for a given point in the $\overline{w}_\text{reh} \times T_\text{reh}$ parameter space, we obtain a family of gravitational-wave spectra, each corresponding to a specific choice of $w_\text{sd}$, as illustrated in figs.~\ref{fig:broken} and \ref{fig:rad}. This degeneracy has important consequences for both the $\Delta N_\text{eff}$ constraint and the detectability of these primordial gravitational waves.

Every mode from $f_\text{end}$ down to $f_\text{sd}$ enters the Hubble sphere during the first epoch---either matter domination ($w_\text{md}=0$) or radiation domination ($w_\text{rd}=1/3$)---, whereas modes with frequencies between $f_\text{sd}$ and $f_\text{reh}$ re-enter during the stiff era. From the Friedmann equation (recall $a(t)\propto t^{2/3(1+w)}\therefore k=aH\propto t^{-(1+3w)/3(1+w)}$) and the definition of the number of $e$-folds as the natural logarithm of the scale factor ratio, the duration of the first epoch ($N_\text{1}$) and of the subsequent stiff-dominated epoch ($N_\text{sd}$) are: \begin{equation}\begin{split}
    N_\text{1} &= -\frac{2}{1+ 3w_\text{1}} \ln \left(\frac{f_\text{sd}}{f_\text{end}}\right),\\
    N_\text{sd} &= -\frac{2}{1 + 3w_\text{sd}} \ln \left(\frac{f_\text{reh}}{f_\text{sd}}\right)\;,
    \label{efoldstwo}
\end{split}\end{equation}
where $w_1 \in \{w_\text{m}, w_\text{r}\}$ denotes the EoS parameter of the first epoch. Furthermore, the total duration of reheating remains constrained by equation~\eqref{eq:ratio2}, yielding
\begin{equation}
    N_\text{T}=N_\text{sd} + N_\text{1} = -\frac{1}{3(1 + \overline{w}_\text{reh})}\ln \left(\frac{\rho_\text{reh}}{\rho_\text{end}}\right)\; ,
    \label{totalefoldstwo}
\end{equation}
where the average EoS for the entire two-stage reheating is now given by (cf. eq.~\eqref{wbar})
\begin{equation}
    \overline{w}_\text{reh} = \frac{w_\text{sd}N_\text{sd} + w_\text{1} N_\text{1}}{N_\text{T}}\; .
    \label{average}
\end{equation}

By combining eqs.~\eqref{efoldstwo}--\eqref{average}, one can analytically solve the system for the transition frequency, $f_\text{sd}$, which yields
\begin{equation}
    f_\text{sd} = f_\text{end} \exp \left[ - \frac{N_\text{T}}{2} \left( \frac{w_\text{sd} - \overline{w}_\text{reh}}{w_\text{sd} - w_1} \right) (1 + 3w_1) \right]\;.
    \label{eq:fsd_exact}
\end{equation}

\section{Results}
\label{sec:results}

\begin{figure}
    \centering
    \begin{picture}(300,180)
        \put(0,0){\includegraphics[width=0.48\textwidth]{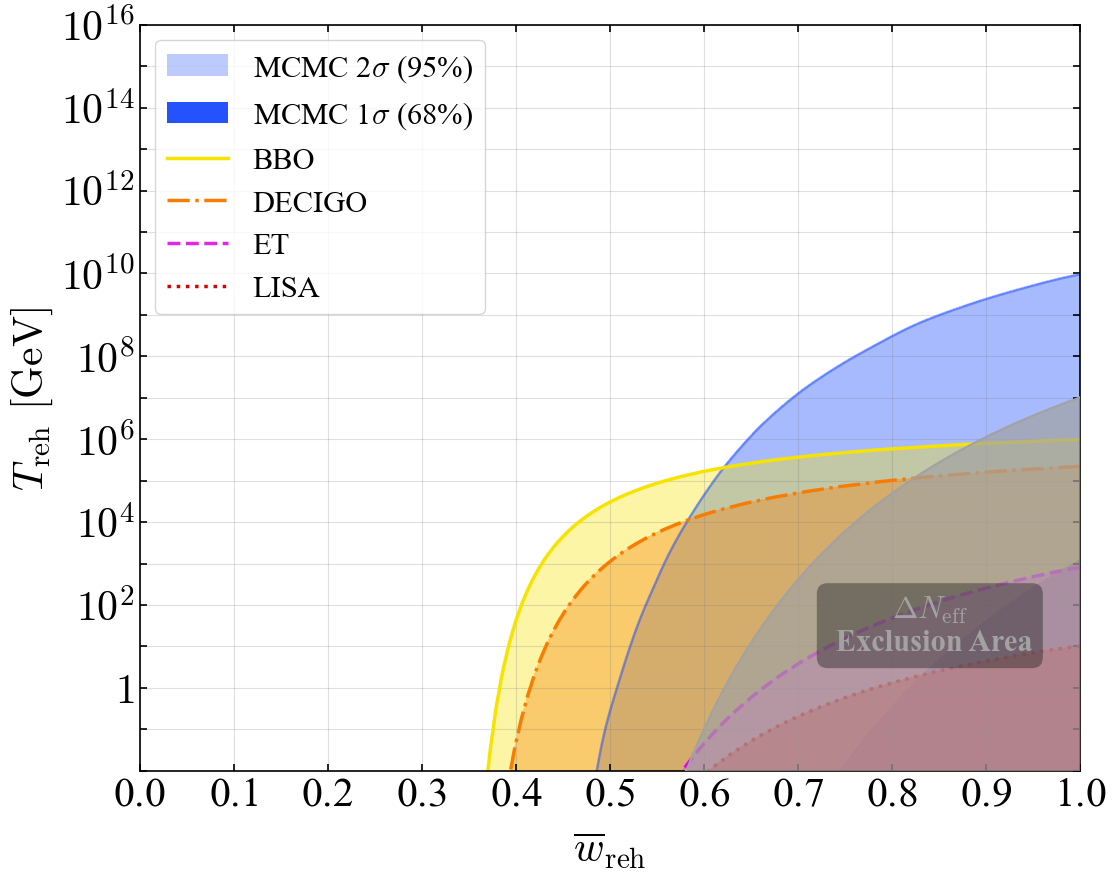}}
        \put(128,158){\fcolorbox{white}{white}{\normalsize{Constant $w_\text{reh}$}}}
    \end{picture}
    \caption{The blue curve represents the $1\sigma$ and $2\sigma$ regions in the $(\overline{w}_\text{reh},T_\text{reh})$ parameter space agreeing with ACT DR6 data for the Starobinsky model. Regions in yellow (BBO), orange (DECIGO), pink (Einstein Telescope), and red (LISA) indicate where $\mathrm{SNR}>10$ for each of these experiments, assuming a reheating phase characterized purely by stiff domination. Finally, the grey curve represents the portion of parameter space excluded by the $\Delta N_\text{eff}$ bound.}
    \label{fig:only sd}
\end{figure}

We first show in fig.~\ref{fig:only sd} the results for the case of a purely stiff-dominated reheating with a constant EoS parameter $w_\text{reh}>1/3$. As seen in fig.~\ref{fig:triangle}, ACT DR6 data pushes this $w_\text{reh}$ towards larger values and relatively low reheating temperatures ($w_\text{reh}\gtrsim 0.7$ and $T_\text{reh}\lesssim 10^{4}$~GeV at $1\sigma$), which in turn are expected to violate $\Delta N_\text{eff}$ bounds. Indeed, fig.~\ref{fig:only sd} shows that, for Starobinsky inflation, the entire region agreeing with ACT DR6 at $1\sigma$ is excluded by the constraint from BBN---a result which is in full agreement with ref.~\cite{Haque:2025uis}.

Part of the $2\sigma$ region allowed by CMB+BAO data evades the $\Delta N_\text{eff}$ bound, but would be within reach of future GW interferometers. This is especially true of DECIGO and BBO, which can probe regions of $\overline{w}_\text{reh}\gtrsim 0.35$ and $T_\text{reh}\lesssim 10^{6}-10^{7}$~GeV. On the other hand, third-generation observatories such as LISA and ET would essentially probe only the region already excluded by BBN. 

\begin{figure}
    \centering
    \begin{picture}(300,180)
        \put(0,0){\includegraphics[width=0.48\textwidth]{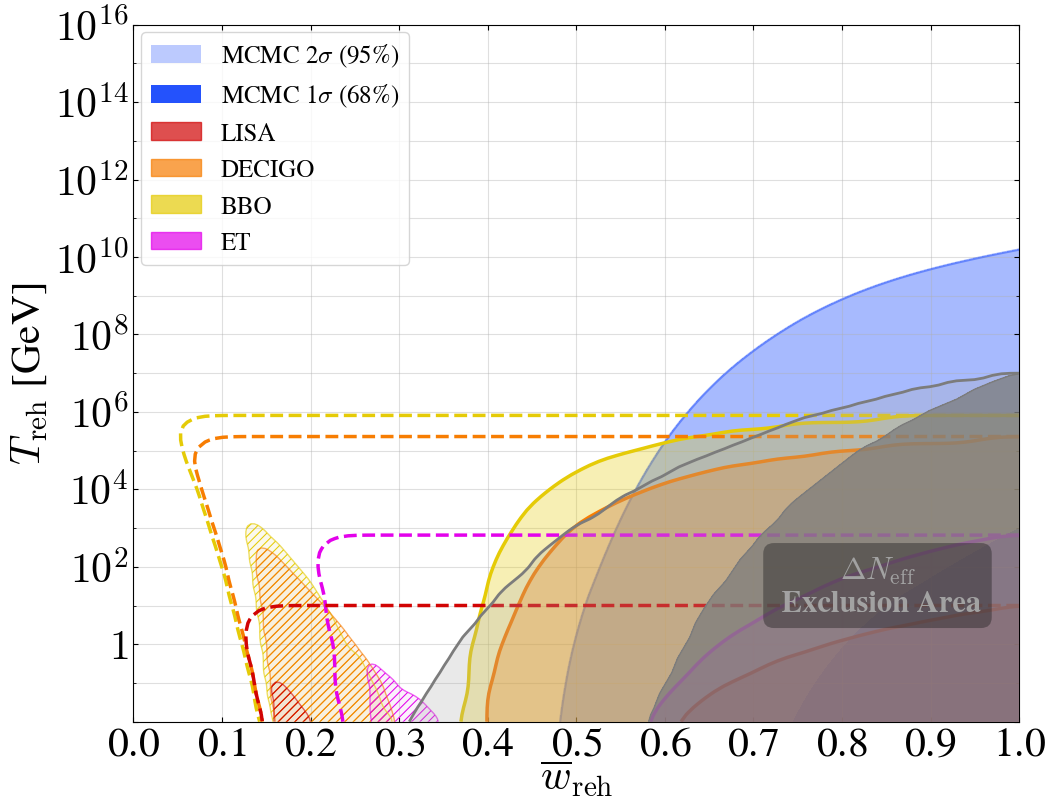}}
        \put(110,145){\fcolorbox{white}{white}{\parbox{2.5cm}{\centering \normalsize Matter $+$ stiff\\ domination}}}
    \end{picture}
    \caption{Similar to fig.~\ref{fig:only sd}, but for a broken-power-law spectrum with an early matter-dominated era. The shape of the spectrum, and hence the detectability and exclusion regions,  now depend not only on $(\overline{w}_\text{reh}, T_\text{reh})$ but also on an additional parameter $w_\text{sd}$. Dashed lines delineate regions of \emph{potential} detectability for BBO (yellow), DECIGO (orange), ET (pink), and LISA (red), in which only a subset of the viable $w_\text{sd}$ values render the gravitational-wave signal detectable. Solid curves enclose regions where SNR~$>10$ for all $w_{\mathrm{sd}}$. Similarly, the outer and inner grey curves define the boundaries of the potentially and certainly excluded regions due to $\Delta N_\text{eff}$ bounds. Finally, hatched areas indicate where $\Omega_{\text{gw}}(f_{\mathrm{sd}})$ lies within the detector's sensitivity, making it possible to probe the broken-power-law feature. Outside these hatched regions, detectors see the GW spectrum as essentially purely stiff, and one can recycle the plot in fig.~\ref{fig:only sd} to read whether a spectrum will be detectable by simply replacing $\overline{w}_\text{reh}\to w_\text{sd}$.}
    \label{fig:sd and md}
\end{figure}

\begin{figure}
    \centering
    \begin{picture}(300,180)
        \put(0,0){\includegraphics[width=0.48\textwidth]{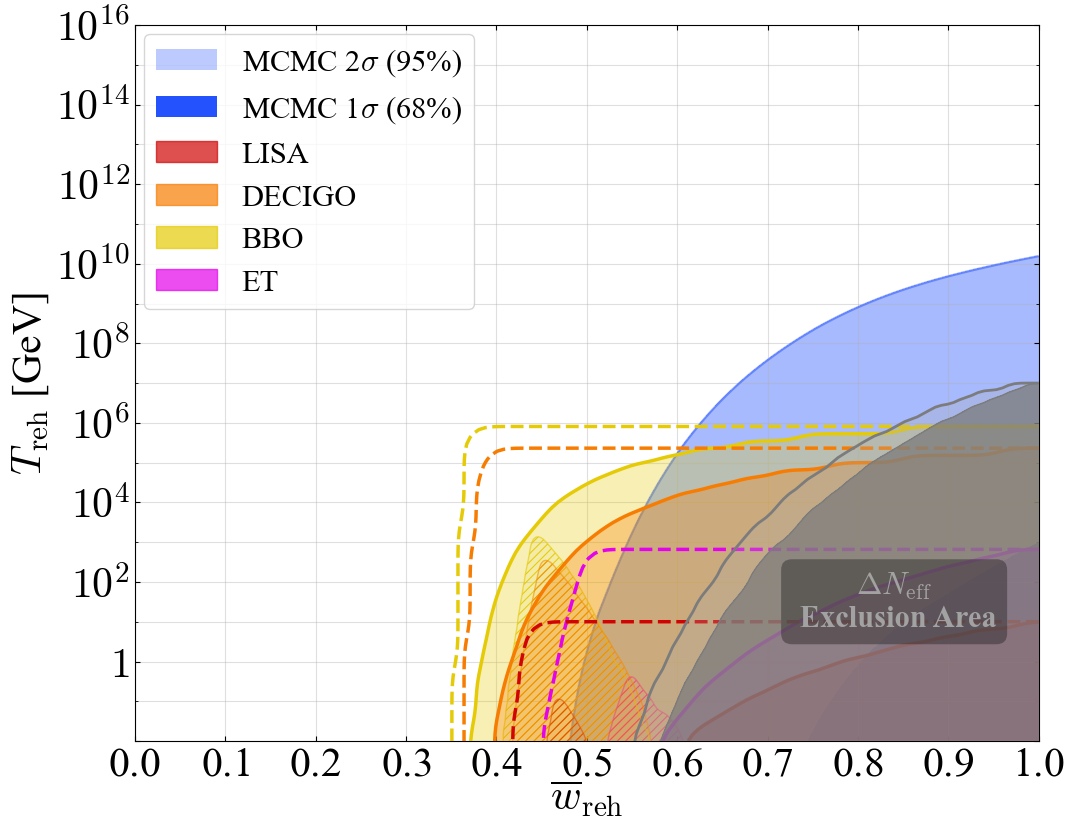}}
        \put(110,145){\fcolorbox{white}{white}{\parbox{3cm}{\centering \normalsize Radiation $+$ stiff \\ domination}}}
    \end{picture}
    \caption{Same as fig.~\ref{fig:sd and md}, but for an early radiation-dominated era preceding the stiff epoch. This modification introduces important structural differences in parameter space. The excluded region closely resembles that of fig.~\ref{fig:only sd}, leaving only a small area under partial exclusion. The hatched areas shift closer to the MCMC analysis results, falling entirely within the total detectability regions of BBO and DECIGO. Finally, the region where the model is partially detectable in gravitational waves while being consistent with CMB+BAO data and not violating the $\Delta N_{\text{eff}}$ bound is substantially expanded when compared to the case in fig. \ref{fig:sd and md}.}     \label{fig:rd and sd}
\end{figure}

Next, we analyze the cases featuring a broken-power-law spectrum. The results for an early matter-dominated phase preceding stiff domination are displayed in fig.~\ref{fig:sd and md}, whereas the scenario with an early radiation-dominated phase followed by a stiff era is shown in fig.~\ref{fig:rd and sd}. Since ACT is only sensitive to the average EoS $\overline{w}_\text{reh}$, the preferred $1\sigma$ and $2\sigma$ regions remain unaltered. However, BBN bounds and detection from interferometers vary with the frequency dependence of the spectrum, and are therefore sensitive to the evolution of the EoS during reheating.

Although one might na\"ively expect that a broken spectrum could help alleviate the $\Delta N_\text{eff}$ bounds, figs.~\ref{fig:sd and md} and \ref{fig:rd and sd} actually demonstrate the opposite. Because the gravitational-wave spectrum now depends on an additional parameter $w_\text{sd}$ (the EoS of the stiff epoch), there is a family of spectra associated to each point $(\overline{w}_\text{reh}, T_\text{reh})$ in parameter space, spanned by the range $\overline{w}_\text{reh} \leq w_\text{sd} \leq 1$, as illustrated in figs.~\ref{fig:broken} and \ref{fig:rad}. The minimum amplitude occurs when $w_\text{sd}=\overline{w}_\text{reh}$ (and $f_\text{sd}=f_\text{end}$), which corresponds to the purely stiff scenario. Being the lowest-amplitude configuration, it sets the minimum threshold for exclusion: if this specific spectrum is ruled out, all other configurations within the family are strictly excluded for that same point in parameter space. This behavior applies equally to both broken spectrum scenarios. Consequently, the fully excluded regions in figs.~\ref{fig:sd and md} and \ref{fig:rd and sd} match the exclusion area of fig.~\ref{fig:only sd}.

But now there are certain combinations $(\overline{w}_\text{reh},T_\text{reh})$ which may or may not be ruled out by $\Delta N_\text{eff}$ bounds, and may or may not be detectable by future GW experiments, depending on the EoS during the stiff epoch, $w_\text{sd}$. We plot the \emph{potentially excluded} region as a light grey band in fig.~\ref{fig:sd and md}, whereas the \emph{potentially detectable} regions are the dashed curves. Since the EoS for matter is $w_\mathrm{m}=0$, in order to have a sufficiently large $\overline{w}_\text{reh}$, an even larger $w_\text{sd}$ is required, meaning a more pronounced blue-tilting. The curves then enter the detectability regions of the interferometers (and also cross the $\Delta N_\text{eff}$ bound) for even lower average values $\overline{w}_\text{reh}$ when compared to the other two scenarios. This physical contrast is readily visible by comparing the spectral amplitudes for the case with $\overline{w}_\text{reh}=0.6$ and $T_\text{reh}=100~\text{GeV}$ in figs.~\ref{fig:broken} and \ref{fig:rad}.

Because the detailed spectrum now depends on an additional parameter not shown in the $(\overline{w}_\text{reh}, T_\text{reh})$ plane, one might wonder how to extract information on whether a certain spectrum (characterized by an EoS $w_\text{sd}$ during the stiff epoch) is detectable or not just by looking at figs.~\ref{fig:sd and md} and \ref{fig:rd and sd}. For this purpose we show in these plots the hashed regions where the amplitude at $f_\text{sd}$\footnote{$f_\text{sd}$ is the frequency when the matter or radiation domination transitions to stiff domination. Note that fixing $w_\text{sd}$ fixes $f_\text{sd}$ as well, cf. eq.~\eqref{eq:fsd_exact}.} is within the sensitivity range of the detectors. If so, this means the detector is able to probe the broken-power-law behaviour, and hence potentially probe details of the reheating history. If not, the interferometer sees the spectrum as if it were purely stiff, in which case one can recycle the plot in fig.~\ref{fig:only sd} and replace $\overline{w}_\text{reh}\to w_\text{sd}$.

Figs.~\ref{fig:only sd}, \ref{fig:sd and md}, and \ref{fig:rd and sd} show that the regions of certain detectability for LISA and ET lie entirely in the region where $\Delta N_\text{eff}$ bounds are violated. For a two-stepped reheating history, the potentially detectable region by these interferometers is somewhat larger, encompassing part of the still allowed $2\sigma$ region. This shows how these third-generation interferometers can be used to probe the history of reheating in a Starobinsky scenario of inflation. In particular, LISA and ET would be able to probe reheating temperatures of up to $T_\text{reh}\sim 10$~GeV and $T_\text{reh}\sim 10^3$~GeV, respectively. For an even farther future, DECIGO and BBO could probe the $2\sigma$ region even in the purely stiff case, testing reheating temperatures of up to $\sim 10^{5}-10^6$~GeV. 

Interestingly, the figures show that LISA and ET could probe a reheating with $\overline{w}_\text{reh}\gtrsim 0.4$ (for a two-stepped reheating with early radiation domination) or $\overline{w}_\text{reh}\gtrsim 0.15-0.25$ (for early matter domination). Likewise, DECIGO and BBO are sensitive to $\overline{w}_\text{reh}\gtrsim 0.35$ or even down to $\overline{w}_\text{reh}\gtrsim 0.15$ for early matter domination. Even though much of these regions are outside the $2\sigma$ region of the P-ACT-LB2-BK18 dataset, we should keep in mind the tension between the CMB and BAO datasets~\cite{SPT-3G:2025bzu, Ferreira:2025lrd} which, if ever alleviated, could enhance the size or location of the $2\sigma$ regions in the $(\overline{w}_\text{reh},T_\text{reh})$ space, thus enhancing the discerning power of third- and later-generation GW experiments.

\subsection*{Can reheating reconcile Starobinsky with CMB+BAO results after all?}

Our findings show that, regardless of the details of the reheating history, there will always be a region of certain exclusion by $\Delta N_\text{eff}$ bounds (the dark grey regions in figs.~\ref{fig:only sd}, \ref{fig:sd and md}, and \ref{fig:rd and sd}), which is larger than the $1\sigma$ region allowed by ACT. This allows us to unambiguously filter the $(\overline{w}_\text{reh} \times T_\text{reh})$ plane to remove these points and re-evaluate the allowed intervals for the interesting parameters.

\begin{figure}
    \centering
    \includegraphics[width=1.0\linewidth]{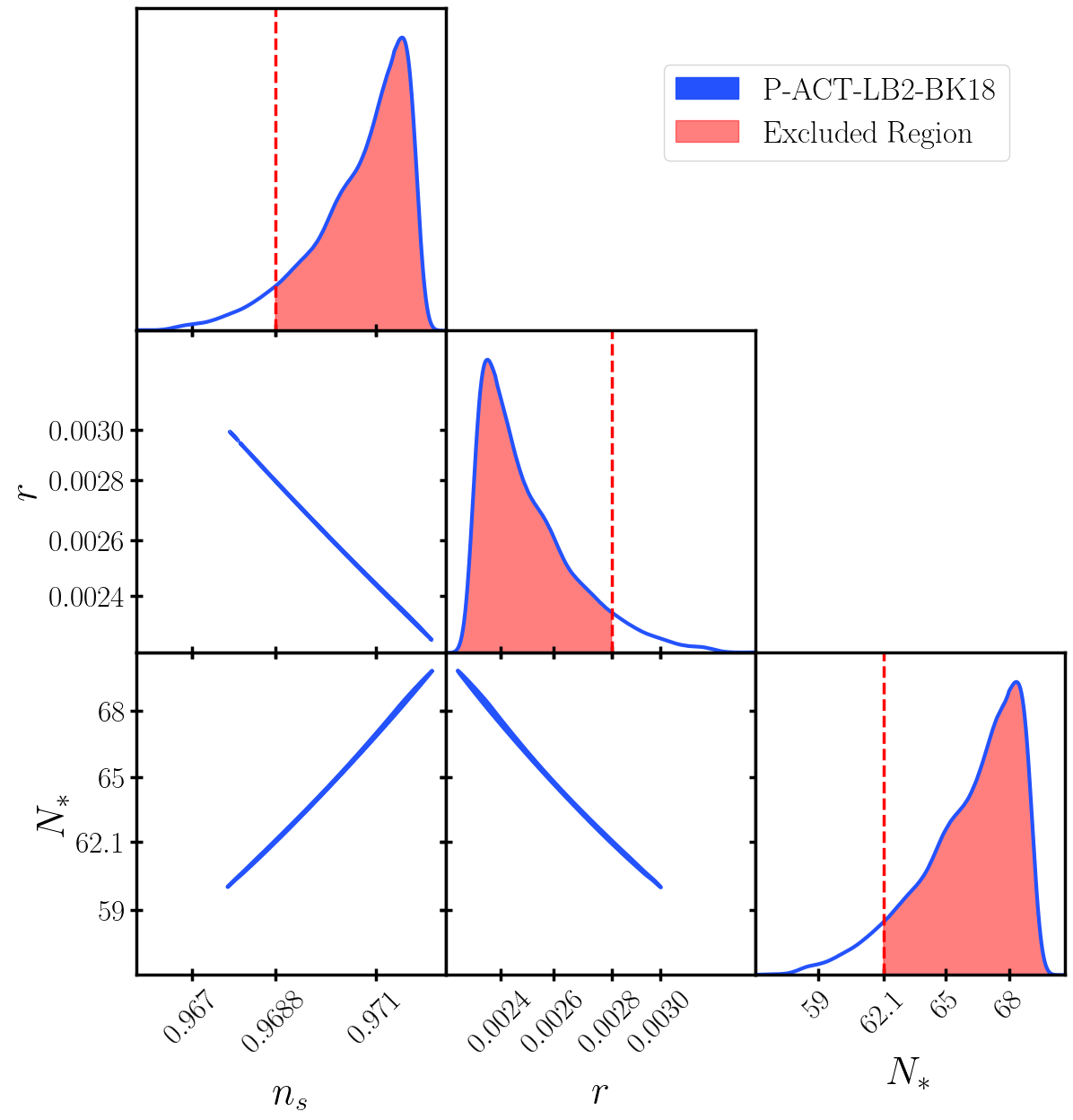}
    \caption{Triangle plot of the $n_s$, $r$, and $N_*$ parameters and their marginalized posterior with the region excluded by the $\Delta N_\text{eff}$ upper bound.}
    \label{fig:ns_excluding_BBN}
\end{figure}

The impact of applying this filter on the observables is shown in fig.~\ref{fig:ns_excluding_BBN}. We find that the amplitude of the primordial scalar power spectrum, $A_s$, remains unaffected by the BBN bounds. However, the other observables are now strictly constrained to $n_s \leq 0.9688$ and $10^3r \geq 2.816$, representing a $2.1\sigma$ tension with the standard model inference derived from the P-ACT-LB2-BK18 dataset ($n_s = 0.9752 \pm 0.0030$).

The appearance of these stringent bounds can be understood by noting that the rescaled reheating parameter, $R_\text{reh}$, is defined in terms of the physical reheating parameters and, indirectly, $\ln V_0$ (see eq.~\ref{eq:Rreh}). Because the reheating parameters drive the largest variations in $R_\text{reh}$, enforcing the $\Delta N_\text{eff}$ upper bound consequently limits the rescaled parameter to $\ln R_\text{reh} \leq -1.324$, which in turn restricts the number of $e$-folds to $N_* \leq 62.09$ via eq.~\ref{eq:Nstar_raw}.

Therefore, while the inclusion of reheating is capable of increasing the number of inflationary $e$-folds and raise the spectral index when compared to the usual interval $N_* = 50-60$, the tension with the data persists at $> 2\sigma$.

\section{Conclusion}
\label{sec:conclusion}

In this paper we performed a Bayesian analysis of reheating in a Starobinsky scenario of inflation to determine the favoured values of the reheating temperature $T_\text{reh}$ and the average EoS parameter during reheating, $\overline{w}_\text{reh}$, in light of the recent combined data from Planck 2018 + ACT DR6 + DESI DR2 + BICEP/Keck 2018. We have taken into account single-stepped and two-stepped reheating histories, both involving a stiff-like fluid at some stage. This stiff epoch helps increase the number of $e$-folds $N_*$ between the first Hubble crossing of the pivot scale $k_*=0.05$~Mpc$^{-1}$ and the end of reheating, which in turn increases the scalar spectral index $n_s$ to larger values than the usual expectation stemming from $N_*=50-60$, possibly placing the model inside the most favourable region in light of the most recent ACT results. Indeed we find that the inclusion of a stiff fluid during reheating successfully raises the spectral index value to $n_s = 0.9706^{+0.0013}_{-0.00043}$ and decreases the tension with the joint data down to $1.4\sigma$, but at the cost of a highly stiff average EoS during reheating. 

We then show that this stiff epoch induces a blue-tilting of the primordial inflationary GWs, which could have observational implications. First, if the blue-tilting is too severe and starts at a low-enough frequency (meaning a low-enough reheating temperature $T_\text{reh}$) then the density of GWs could be large enough to violate known bounds in the amount of radiation during BBN (so-called $\Delta N_\text{eff}$ bounds). An important result of this paper is that this constraint eliminates the region where Starobinsky's model agrees with ACT at $1\sigma$ C.L. This is in agreement with recent results from ref.~\cite{Haque:2025uis}, but here we show that this holds even for more complicated reheating histories, e.g. involving multi-stepped epochs. 

Specifically, we first analyzed the case where reheating first proceeds with a matter-dominated fluid and later decays into stiff-like matter. This is the most realistic scenario for reheating in Starobinsky inflation, since a matter-dominated epoch is naturally induced by the oscillations of the inflaton around a quadratic well~\cite{Dorsch:2024nan, Dorsch:2026ref}. Na\"ively, one would expect that the bounds would be less stringent because the matter-dominated epoch would lead to red-tilting, and so it would seem less likely that the GW spectrum could grow to values that are too large. However, since the EoS parameter during matter domination is $w_\text{m}=0$, in order to have a large average $\overline{w}_\text{reh}$ one needs an even larger stiff EoS, and therefore a more blue-tilted spectrum. The less extreme blue-tilting occurs when the spectrum never breaks (i.e. never red-tilts) up to the largest allowed frequency $f_\text{end}$---but this corresponds precisely to the situation analyzed in the ``pure stiff'' case described previously, which excludes the $1\sigma$ region. 

Recognizing that the problem becomes more severe due to $w_\text{m}=0$ (thus pushing the EoS of the stiff epoch to even larger values in order to have a sufficiently high $\overline{w}_\text{reh}$ as required by CMB+BAO results), we also analyzed the case when reheating proceeds first by radiation domination followed by stiff-fluid domination. We have indeed checked that now the blue-tilting needs to be less severe, but again the mildest case is when the spectrum is solely blue-tilted up to $f_\text{end}$. This is indeed a general trend, and we therefore conclude that \emph{even considering the effects of a stiff-dominated reheating, the tension between Starobinsky's prediction of $n_s$ and the recent ACT DR6 results cannot be brought to less than $1\sigma$}. 

What about the $2\sigma$ region? Part of it is still allowed by the BBN bounds. Now, the blue-tilting also has the effect of potentially placing the primordial GW spectra inside the sensitivity range of near-future detectors, and it would be interesting to know whether they could be used to extract information on the detailed history of reheating. Our analysis showed that, for a reheating dominated entirely by a stiff fluid, the third-generation observatories (LISA and Einstein Telescope) can only access regions already excluded by BBN. However DECIGO and BBO would be able to test the non-excluded region allowed by ACT to $2\sigma$ up to reheating temperatures $T_\text{reh}\lesssim 10^5$~GeV. For a two-stepped reheating (with early matter  or radiation domination) even the third-generation experiments could be sensitive to the $2\sigma$ region, and in fact to even much lower values of $\overline{w}_\text{reh}\gtrsim 0.15-0.25$ and $T_\text{reh}\lesssim 10-10^3$~GeV (for early matter domination). Much of this region is in tension with CMB+BAO results, but it is nevertheless interesting to know that GW interferometers could probe them, especially taking into account the known tensions in the CMB+BAO datasets that led to the ACT results~\cite{ACTDR6, Ferreira:2025lrd}. For an earlier radiation-dominated epoch, LISA and ET could probe the allowed $2\sigma$ region as well, exploring values of $\overline{w}_\text{reh}\gtrsim 0.4-0.45$ and $T_\text{reh}\lesssim 10-10^3$~GeV. DECIGO and BBO would of course be sensitive to even larger regions.

Overall, gravitational-wave experiments will enable us to test multiple reheating scenarios that are favoured by current CMB+BAO analyses. Should no signal be found in these experiments, many reheating scenarios will be definitely excluded, and the model will be cornered down to smaller regions of parameter space. Still, it is interesting to note that there is a region of $T_\text{reh}\in (10^7, 10^{10})$~GeV where the tension with ACT is less than $3\sigma$ and these stiff scenarios are neither excluded by BBN nor detectable by planned interferometers. This is all the more remarkable when one notices that Starobinsky inflation with purely gravitational reheating predicts a reheating temperature of $T_\text{reh}\sim 10^9$~GeV, thus within this range~\cite{Dorsch:2024nan, Dorsch:2026ref}.

\section*{Acknowledgements}
GCD would like to thank CNPq for financial support under grant no. 307565/2025-4. LPCL and LCAM acknowledge the financial support of the CAPES agency. BMDS greatly thanks CNPq for the financial support. LMBA is grateful for the support of the MIT Department of Physics.

\appendix
\section{CLASS Settings}

Below we present the \texttt{CLASS} (v3.3.4) input configuration, following the input and precision parameters from the ACT DR6 analysis on extended models~\cite{ACTDR6}. We fixed the non-inflationary cosmological parameters from the $\Lambda$CDM model to the mean values found in P-ACT-LB2 joint analysis. Sampling them together with the Starobinsky inflation plus reheating did not yield significant deviations on the $\ln{V_0}\times \ln{R_\text{reh}}$ plane.

\begin{lstlisting}[caption={CLASS configuration and high-precision parameters.}, label={lst:class_params}]
h: 0.6843,
omega_b: 0.02258,
omega_cdm: 0.1174,
tau_reio: 0.0643,
k_pivot = 0.05
N_ncdm = 1
m_ncdm = 0.06
N_ur = 2.0308
T_cmb = 2.7255
YHe = BBN
non_linear = hmcode
hmcode_version = 2020
hmcode_max_k_extra = 0
output = lCl tCl pCl mPk
modes = s, t
lensing = yes
recombination = HyRec
accurate_lensing = 1
perturbations_sampling_stepsize = 0.05
l_logstep = 1.025
l_linstep = 20
l_max_scalars = 9500          
P_k_max_h/Mpc = 100.0
delta_l_max = 1800
l_switch_limber = 30.0
hyper_sampling_flat = 32.0
l_max_g = 40
l_max_ur = 35
l_max_pol_g = 60
ur_fluid_approximation = 2
ur_fluid_trigger_tau_over_tau_k = 130.0
radiation_streaming_approximation = 2
radiation_streaming_trigger_tau_over_tau_k = 240.0
hyper_flat_approximation_nu = 7000.0
transfer_neglect_delta_k_S_t0 = 0.17
transfer_neglect_delta_k_S_t1 = 0.05
transfer_neglect_delta_k_S_t2 = 0.17
transfer_neglect_delta_k_S_e = 0.17
start_small_k_at_tau_c_over_tau_h = 0.0004
start_large_k_at_tau_h_over_tau_k = 0.05
tight_coupling_trigger_tau_c_over_tau_h = 0.005
tight_coupling_trigger_tau_c_over_tau_k = 0.008
start_sources_at_tau_c_over_tau_h = 0.006
l_max_ncdm = 30
tol_ncdm_synchronous = 1e-6
\end{lstlisting}

\bibliographystyle{JHEP}
\bibliography{Refs}

@article{Maggiore:1999vm,
    author = "Maggiore, Michele",
    title = "{Gravitational wave experiments and early universe cosmology}",
    eprint = "gr-qc/9909001",
    archivePrefix = "arXiv",
    reportNumber = "IFUP-TH-20-99",
    doi = "10.1016/S0370-1573(99)00102-7",
    journal = "Phys. Rept.",
    volume = "331",
    pages = "283--367",
    year = "2000"
}

@book{Maggiore:2018sht,
    author = "Maggiore, Michele",
    title = "{Gravitational Waves. Vol. 2: Astrophysics and Cosmology}",
    isbn = "978-0-19-857089-9",
    publisher = "Oxford University Press",
    year = "2018"
}

@misc{alves2026gravitationalwavesbigbang,
      title={Gravitational Waves from the Big Bang}, 
      author={Lucas Martins Barreto Alves},
      year={2026},
      eprint={2511.17659},
      archivePrefix={arXiv},
      primaryClass={gr-qc},
      url={https://arxiv.org/abs/2511.17659}, 
}

@book{Misner:1973prb,
    author = "Misner, Charles W. and Thorne, K. S. and Wheeler, J. A.",
    title = "{Gravitation}",
    isbn = "978-0-7167-0344-0, 978-0-691-17779-3",
    publisher = "W. H. Freeman",
    address = "San Francisco",
    year = "1973"
}

@book{Kolb:1990vq,
    author = "Kolb, Edward W.",
    title = "{The Early Universe}",
    reportNumber = "FERMILAB-BOOK-1990-01",
    doi = "10.1201/9780429492860",
    isbn = "978-0-429-49286-0, 978-0-201-62674-2",
    publisher = "Taylor and Francis",
    volume = "69",
    month = "5",
    year = "2019"
}

@article{Mossa:2020gjc,
    author = "Mossa, V. and others",
    title = "{The baryon density of the Universe from an improved rate of deuterium burning}",
    doi = "10.1038/s41586-020-2878-4",
    journal = "Nature",
    volume = "587",
    number = "7833",
    pages = "210--213",
    year = "2020"
}

@article{Hsyu:2020uqb,
    author = "Hsyu, Tiffany and Cooke, Ryan J. and Prochaska, J. Xavier and Bolte, Michael",
    title = "{The PHLEK Survey: A New Determination of the Primordial Helium Abundance}",
    eprint = "2005.12290",
    archivePrefix = "arXiv",
    primaryClass = "astro-ph.GA",
    doi = "10.3847/1538-4357/ab91af",
    journal = "Astrophys. J.",
    volume = "896",
    number = "1",
    pages = "77",
    year = "2020"
}

@article{Cooke:2017cwo,
    author = "Cooke, Ryan J. and Pettini, Max and Steidel, Charles C.",
    title = "{One Percent Determination of the Primordial Deuterium Abundance}",
    eprint = "1710.11129",
    archivePrefix = "arXiv",
    primaryClass = "astro-ph.CO",
    doi = "10.3847/1538-4357/aaab53",
    journal = "Astrophys. J.",
    volume = "855",
    number = "2",
    pages = "102",
    year = "2018"
}

@article{Cyburt:2015mya,
    author = "Cyburt, Richard H. and Fields, Brian D. and Olive, Keith A. and Yeh, Tsung-Han",
    title = "{Big Bang Nucleosynthesis: 2015}",
    eprint = "1505.01076",
    archivePrefix = "arXiv",
    primaryClass = "astro-ph.CO",
    reportNumber = "UMN-TH-3432-15, FTPI-MINN-15-19",
    doi = "10.1103/RevModPhys.88.015004",
    journal = "Rev. Mod. Phys.",
    volume = "88",
    pages = "015004",
    year = "2016"
}

@misc{duval2024investigatingcosmichistoriesstiff,
      title={Investigating cosmic histories with a stiff era through Gravitational Waves}, 
      author={Hannah Duval and Sachiko Kuroyanagi and Alberto Mariotti and Alba Romero-Rodríguez and Mairi Sakellariadou},
      year={2024},
      eprint={2405.10201},
      archivePrefix={arXiv},
      primaryClass={gr-qc},
      url={https://arxiv.org/abs/2405.10201}, 
}

@article{Boyle:2007zx,
    author = "Boyle, Latham A. and Buonanno, Alessandra",
    title = "{Relating gravitational wave constraints from primordial nucleosynthesis, pulsar timing, laser interferometers, and the CMB: Implications for the early Universe}",
    eprint = "0708.2279",
    archivePrefix = "arXiv",
    primaryClass = "astro-ph",
    doi = "10.1103/PhysRevD.78.043531",
    journal = "Phys. Rev. D",
    volume = "78",
    pages = "043531",
    year = "2008"
}

@article{Torrado_2021,
doi = {10.1088/1475-7516/2021/05/057},
url = {https://doi.org/10.1088/1475-7516/2021/05/057},
year = {2021},
month = {may},
publisher = {IOP Publishing},
volume = {2021},
number = {05},
pages = {057},
author = {Torrado, Jesús and Lewis, Antony},
title = {Cobaya: code for Bayesian analysis of hierarchical physical models},
journal = {Journal of Cosmology and Astroparticle Physics}
}

@article{Foreman-Mackey_2013,
doi = {10.1086/670067},
url = {https://doi.org/10.1086/670067},
year = {2013},
month = {feb},
publisher = {University of Chicago Press},
volume = {125},
number = {925},
pages = {306},
author = {Foreman-Mackey, Daniel and Hogg, David W. and Lang, Dustin and Goodman, Jonathan},
title = {emcee: The MCMC Hammer},
journal = {Publications of the Astronomical Society of the Pacific}
}

@software{2019ascl_soft10019T,
       author = {{Torrado}, Jes{\'u}s and {Lewis}, Antony},
        title = "{Cobaya: Bayesian analysis in cosmology}",
 howpublished = {Astrophysics Source Code Library, record ascl:1910.019},
         year = 2019,
        month = oct,
          eid = {ascl:1910.019},
archivePrefix = {ascl},
       eprint = {1910.019},
       adsurl = {https://ui.adsabs.harvard.edu/abs/2019ascl.soft10019T}
}

@article{Diego_Blas_2011,
doi = {10.1088/1475-7516/2011/07/034},
url = {https://doi.org/10.1088/1475-7516/2011/07/034},
year = {2011},
month = {jul},
publisher = {},
volume = {2011},
number = {07},
pages = {034},
author = {Diego Blas and Julien Lesgourgues and Thomas Tram},
title = {The Cosmic Linear Anisotropy Solving System (CLASS).
 Part II: Approximation schemes},
journal = {Journal of Cosmology and Astroparticle Physics}
}

@article{Lewis_2025,
doi = {10.1088/1475-7516/2025/08/025},
url = {https://doi.org/10.1088/1475-7516/2025/08/025},
year = {2025},
month = {aug},
publisher = {IOP Publishing},
volume = {2025},
number = {08},
pages = {025},
author = {Lewis, Antony},
title = {GetDist: a Python package for analysing Monte Carlo samples},
journal = {Journal of Cosmology and Astroparticle Physics}
}

@article{km3q-rm34,
  title = {ACT observations, reheating, and Starobinsky and Higgs inflation},
  author = {Zharov, D. S. and Sobol, O. O. and Vilchinskii, S. I.},
  journal = {Phys. Rev. D},
  volume = {112},
  issue = {2},
  pages = {023544},
  numpages = {11},
  year = {2025},
  month = {Jul},
  publisher = {American Physical Society},
  doi = {10.1103/km3q-rm34},
  url = {https://link.aps.org/doi/10.1103/km3q-rm34}
}

@article{Haque:2021dha,
    author = "Haque, Md Riajul and Maity, Debaprasad and Paul, Tanmoy and Sriramkumar, L.",
    title = "{Decoding the phases of early and late time reheating through imprints on primordial gravitational waves}",
    eprint = "2105.09242",
    archivePrefix = "arXiv",
    primaryClass = "astro-ph.CO",
    doi = "10.1103/PhysRevD.104.063513",
    journal = "Phys. Rev. D",
    volume = "104",
    number = "6",
    pages = "063513",
    year = "2021"
}

@article{ACTDR6,
    author = "Calabrese, Erminia and others",
    collaboration = "Atacama Cosmology Telescope",
    title = "{The Atacama Cosmology Telescope: DR6 constraints on extended cosmological models}",
    eprint = "2503.14454",
    archivePrefix = "arXiv",
    primaryClass = "astro-ph.CO",
    reportNumber = "FERMILAB-PUB-25-0157-PPD",
    doi = "10.1088/1475-7516/2025/11/063",
    journal = "JCAP",
    volume = "11",
    pages = "063",
    year = "2025"
}

@article{Fixsen_2009,
doi = {10.1088/0004-637X/707/2/916},
url = {https://doi.org/10.1088/0004-637X/707/2/916},
year = {2009},
month = {nov},
publisher = {The American Astronomical Society},
volume = {707},
number = {2},
pages = {916},
author = {Fixsen, D. J.},
title = {THE TEMPERATURE OF THE COSMIC MICROWAVE BACKGROUND},
journal = {The Astrophysical Journal}
}

@article{Starobinsky:1980te,
    author = "Starobinsky, Alexei A.",
    editor = "Khalatnikov, I. M. and Mineev, V. P.",
    title = "{A New Type of Isotropic Cosmological Models Without Singularity}",
    doi = "10.1016/0370-2693(80)90670-X",
    journal = "Phys. Lett. B",
    volume = "91",
    pages = "99--102",
    year = "1980"
}

@article{Starobinsky:1983zz,
    author = "Starobinsky, A. A.",
    title = "{The Perturbation Spectrum Evolving from a Nonsingular Initially De-Sitter Cosmology and the Microwave Background Anisotropy}",
    journal = "Sov. Astron. Lett.",
    volume = "9",
    pages = "302",
    year = "1983"
}

@article{tHooft:1974toh,
    author = "'t Hooft, Gerard and Veltman, M. J. G.",
    title = "{One-loop divergencies in the theory of gravitation}",
    doi = "10.1142/9789814539395_0001",
    journal = "Ann. Inst. H. Poincare Phys. Theor. A",
    volume = "20",
    number = "1",
    pages = "69--94",
    year = "1974"
}

@article{Donoghue:1994dn,
    author = "Donoghue, John F.",
    title = "{General relativity as an effective field theory: The leading quantum corrections}",
    eprint = "gr-qc/9405057",
    archivePrefix = "arXiv",
    reportNumber = "UMHEP-408",
    doi = "10.1103/PhysRevD.50.3874",
    journal = "Phys. Rev. D",
    volume = "50",
    pages = "3874--3888",
    year = "1994"
}

@article{Dorsch:2024nan,
    author = "Dorsch, Gl{\'a}uber C. and Miranda, Luiz and Yokomizo, Nelson",
    title = "{Gravitational reheating in Starobinsky inflation}",
    eprint = "2406.04161",
    archivePrefix = "arXiv",
    primaryClass = "gr-qc",
    doi = "10.1088/1475-7516/2024/11/050",
    journal = "JCAP",
    volume = "11",
    pages = "050",
    year = "2024"
}

@article{Dorsch:2026ref,
    author = "Dorsch, Gl{\'a}uber C. and Miranda, Luiz Carlos and Yokomizo, Nelson",
    title = "{Reheating after Starobinsky Inflation in the Jordan Frame}",
    eprint = "2603.04497",
    archivePrefix = "arXiv",
    primaryClass = "gr-qc",
    month = "3",
    year = "2026"
}

@article{Ferreira:2025lrd,
    author = "Ferreira, Elisa G. M. and McDonough, Evan and Balkenhol, Lennart and Kallosh, Renata and Knox, Lloyd and Linde, Andrei",
    title = "{BAO-CMB tension and implications for inflation}",
    eprint = "2507.12459",
    archivePrefix = "arXiv",
    primaryClass = "astro-ph.CO",
    doi = "10.1103/lq71-b84v",
    journal = "Phys. Rev. D",
    volume = "113",
    number = "4",
    pages = "043524",
    year = "2026"
}

@article{Liddle:2003as,
    author = "Liddle, Andrew R and Leach, Samuel M",
    title = "{How long before the end of inflation were observable perturbations produced?}",
    eprint = "astro-ph/0305263",
    archivePrefix = "arXiv",
    doi = "10.1103/PhysRevD.68.103503",
    journal = "Phys. Rev. D",
    volume = "68",
    pages = "103503",
    year = "2003"
}

@article{Dodelson:2003vq,
    author = "Dodelson, Scott and Hui, Lam",
    title = "{A Horizon Ratio Bound for Inflationary Fluctuations}",
    eprint = "astro-ph/0305113",
    archivePrefix = "arXiv",
    reportNumber = "FERMILAB-PUB-03-188-A",
    doi = "10.1103/PhysRevLett.91.131301",
    journal = "Phys. Rev. Lett.",
    volume = "91",
    pages = "131301",
    year = "2003"
}

@article{ACTDR6_cosmo_param,
    author = "Louis, Thibaut and others",
    collaboration = "Atacama Cosmology Telescope",
    title = "{The Atacama Cosmology Telescope: DR6 power spectra, likelihoods and {\ensuremath{\Lambda}}CDM parameters}",
    eprint = "2503.14452",
    archivePrefix = "arXiv",
    primaryClass = "astro-ph.CO",
    reportNumber = "FERMILAB-PUB-25-0071-PPD",
    doi = "10.1088/1475-7516/2025/11/062",
    journal = "JCAP",
    volume = "11",
    pages = "062",
    year = "2025"
}

@article{SPT-3G:2025bzu,
    author = "Camphuis, E. and others",
    collaboration = "SPT-3G",
    title = "{SPT-3G D1: CMB temperature and polarization power spectra and cosmology from 2019 and 2020 observations of the SPT-3G main field}",
    eprint = "2506.20707",
    archivePrefix = "arXiv",
    primaryClass = "astro-ph.CO",
    reportNumber = "FERMILAB-PUB-25-0144-PPD",
    doi = "10.1103/7wt3-9v2y",
    journal = "Phys. Rev. D",
    volume = "113",
    number = "8",
    pages = "083504",
    year = "2026"
}

@article{BICEP:2021xfz,
    author = "Ade, P. A. R. and others",
    collaboration = "BICEP, Keck",
    title = "{Improved Constraints on Primordial Gravitational Waves using Planck, WMAP, and BICEP/Keck Observations through the 2018 Observing Season}",
    eprint = "2110.00483",
    archivePrefix = "arXiv",
    primaryClass = "astro-ph.CO",
    doi = "10.1103/PhysRevLett.127.151301",
    journal = "Phys. Rev. Lett.",
    volume = "127",
    number = "15",
    pages = "151301",
    year = "2021"
}

@article{Planck:2018vyg,
    author = "Aghanim, N. and others",
    collaboration = "Planck",
    title = "{Planck 2018 results. VI. Cosmological parameters}",
    eprint = "1807.06209",
    archivePrefix = "arXiv",
    primaryClass = "astro-ph.CO",
    doi = "10.1051/0004-6361/201833910",
    journal = "Astron. Astrophys.",
    volume = "641",
    pages = "A6",
    year = "2020",
    note = "[Erratum: Astron.Astrophys. 652, C4 (2021)]"
}

@article{DESIDR2,
    author = "Abdul Karim, M. and others",
    collaboration = "DESI",
    title = "{DESI DR2 results. II. Measurements of baryon acoustic oscillations and cosmological constraints}",
    eprint = "2503.14738",
    archivePrefix = "arXiv",
    primaryClass = "astro-ph.CO",
    reportNumber = "FERMILAB-PUB-25-0169-PPD",
    doi = "10.1103/tr6y-kpc6",
    journal = "Phys. Rev. D",
    volume = "112",
    number = "8",
    pages = "083515",
    year = "2025"
}

@article{Drees:2025ngb,
    author = "Drees, Manuel and Xu, Yong",
    title = "{Refined predictions for Starobinsky inflation and post-inflationary constraints in light of ACT}",
    eprint = "2504.20757",
    archivePrefix = "arXiv",
    primaryClass = "astro-ph.CO",
    reportNumber = "MITP-25-033",
    doi = "10.1016/j.physletb.2025.139612",
    journal = "Phys. Lett. B",
    volume = "867",
    pages = "139612",
    year = "2025"
}

@article{Zharov:2025zjg,
    author = "Zharov, D. S. and Sobol, O. O. and Vilchinskii, S. I.",
    title = "{ACT observations, reheating, and Starobinsky and Higgs inflation}",
    eprint = "2505.01129",
    archivePrefix = "arXiv",
    primaryClass = "astro-ph.CO",
    doi = "10.1103/km3q-rm34",
    journal = "Phys. Rev. D",
    volume = "112",
    number = "2",
    pages = "023544",
    year = "2025"
}

@article{Haque:2025uis,
    author = "Haque, Md Riajul and Pal, Sourav and Paul, Debarun",
    title = "{Improved predictions on Higgs-Starobinsky inflation and reheating with ACT DR6 and primordial gravitational waves}",
    eprint = "2505.04615",
    archivePrefix = "arXiv",
    primaryClass = "astro-ph.CO",
    doi = "10.1016/j.physletb.2025.139852",
    journal = "Phys. Lett. B",
    volume = "869",
    pages = "139852",
    year = "2025"
}

@article{GRRminus1,
author = {Andrew Gelman and Donald B. Rubin},
title = {Inference from Iterative Simulation Using Multiple Sequences},
volume = {7},
journal = {Statistical Science},
number = {4},
publisher = {Institute of Mathematical Statistics},
pages = {457 -- 472},
year = {1992},
doi = {10.1214/ss/1177011136},
URL = {https://doi.org/10.1214/ss/1177011136}
}

@article{Nakayama:2008wy,
    author = "Nakayama, Kazunori and Saito, Shun and Suwa, Yudai and Yokoyama, Jun'ichi",
    title = "{Probing reheating temperature of the universe with gravitational wave background}",
    eprint = "0804.1827",
    archivePrefix = "arXiv",
    primaryClass = "astro-ph",
    reportNumber = "RESCEU-7-08, UTAP-596",
    doi = "10.1088/1475-7516/2008/06/020",
    journal = "JCAP",
    volume = "06",
    pages = "020",
    year = "2008"
}

@article{Guzzetti:2016mkm,
    author = "Guzzetti, M. C. and Bartolo, N. and Liguori, M. and Matarrese, S.",
    title = "{Gravitational waves from inflation}",
    eprint = "1605.01615",
    archivePrefix = "arXiv",
    primaryClass = "astro-ph.CO",
    doi = "10.1393/ncr/i2016-10127-1",
    journal = "Riv. Nuovo Cim.",
    volume = "39",
    number = "9",
    pages = "399--495",
    year = "2016"
}

@article{Thrane:2013oya,
    author = "Thrane, Eric and Romano, Joseph D.",
    title = "{Sensitivity curves for searches for gravitational-wave backgrounds}",
    eprint = "1310.5300",
    archivePrefix = "arXiv",
    primaryClass = "astro-ph.IM",
    doi = "10.1103/PhysRevD.88.124032",
    journal = "Phys. Rev. D",
    volume = "88",
    number = "12",
    pages = "124032",
    year = "2013"
}

@misc{gwplotter_web,
  author = {Moore, Christopher J. and Cole, Robert H. and Berry, Christopher P. L.},
  title = {{GWplotter}},
  howpublished = {\url{https://gwplotter.com/}},
  year = {2015},
  note = {Accessed: 2026-06-23}
}

@article{Moore_2015,
doi = {10.1088/0264-9381/32/1/015014},
url = {https://doi.org/10.1088/0264-9381/32/1/015014},
year = {2014},
month = {dec},
publisher = {IOP Publishing},
volume = {32},
number = {1},
pages = {015014},
author = {Moore, C J and Cole, R H and Berry, C P L},
title = {Gravitational-wave sensitivity curves},
journal = {Classical and Quantum Gravity}
}

@article{Martin_2006,
doi = {10.1088/1475-7516/2006/08/009},
url = {https://doi.org/10.1088/1475-7516/2006/08/009},
year = {2006},
month = {aug},
publisher = {},
volume = {2006},
number = {08},
pages = {009},
author = {Martin, Jérôme and Ringeval, Christophe},
title = {Inflation after WMAP3: confronting the slow-roll and exact power spectra with CMB
data},
journal = {Journal of Cosmology and Astroparticle Physics},
}

@article{Martin_2014,
doi = {10.1088/1475-7516/2014/03/039},
url = {https://doi.org/10.1088/1475-7516/2014/03/039},
year = {2014},
month = {mar},
publisher = {},
volume = {2014},
number = {03},
pages = {039},
author = {Martin, Jérôme and Ringeval, Christophe and Trotta, Roberto and Vennin, Vincent},
title = {The best inflationary models after Planck},
journal = {Journal of Cosmology and Astroparticle Physics},
}

@article{PhysRevD.70.043506,
  title = {What is the lowest possible reheating temperature?},
  author = {Hannestad, Steen},
  journal = {Phys. Rev. D},
  volume = {70},
  issue = {4},
  pages = {043506},
  numpages = {8},
  year = {2004},
  month = {Aug},
  publisher = {American Physical Society},
  doi = {10.1103/PhysRevD.70.043506},
  url = {https://link.aps.org/doi/10.1103/PhysRevD.70.043506}
}

@article{10.1145/3338517,
author = {Cartis, Coralia and Fiala, Jan and Marteau, Benjamin and Roberts, Lindon},
title = {Improving the Flexibility and Robustness of Model-based Derivative-free Optimization Solvers},
year = {2019},
issue_date = {September 2019},
publisher = {Association for Computing Machinery},
address = {New York, NY, USA},
volume = {45},
number = {3},
issn = {0098-3500},
url = {https://doi.org/10.1145/3338517},
doi = {10.1145/3338517},
month = aug,
articleno = {32},
numpages = {41}
}

@article{DeFelice:2010aj,
    author = "De Felice, Antonio and Tsujikawa, Shinji",
    title = "{f(R) theories}",
    eprint = "1002.4928",
    archivePrefix = "arXiv",
    primaryClass = "gr-qc",
    doi = "10.12942/lrr-2010-3",
    journal = "Living Rev. Rel.",
    volume = "13",
    pages = "3",
    year = "2010"
}
\end{document}